\documentclass[12pt]{iopart}
\usepackage{graphicx}

\begin{document}
\bibliographystyle{plain}

\title{Cataclysmic variables}

\author{Robert Connon Smith}

\address{Department of Physics and Astronomy, University of Sussex, Falmer,
Brighton BN1 9QH, UK}

\ead{r.c.smith@sussex.ac.uk}

\begin{abstract}Cataclysmic variables are binary stars in which a relatively normal star is transferring mass to its compact companion. This interaction gives rise to a rich range of behaviour, of which the most noticeable are the outbursts that give the class its name. Novae belong to the class, as do the less well known dwarf novae and magnetic systems. Novae draw their energy from nuclear reactions, while dwarf novae rely on gravity to power their smaller eruptions. All the different classes of cataclysmic variable can be accommodated within a single framework and this article will describe the framework, review the properties of the main types of system and discuss models of the outbursts and of the long-term evolution.
\end{abstract}

\maketitle

\section{Introduction}\label{intro}

\begin{figure}[h]
\begin{center}
\includegraphics[width=8cm,angle=270]{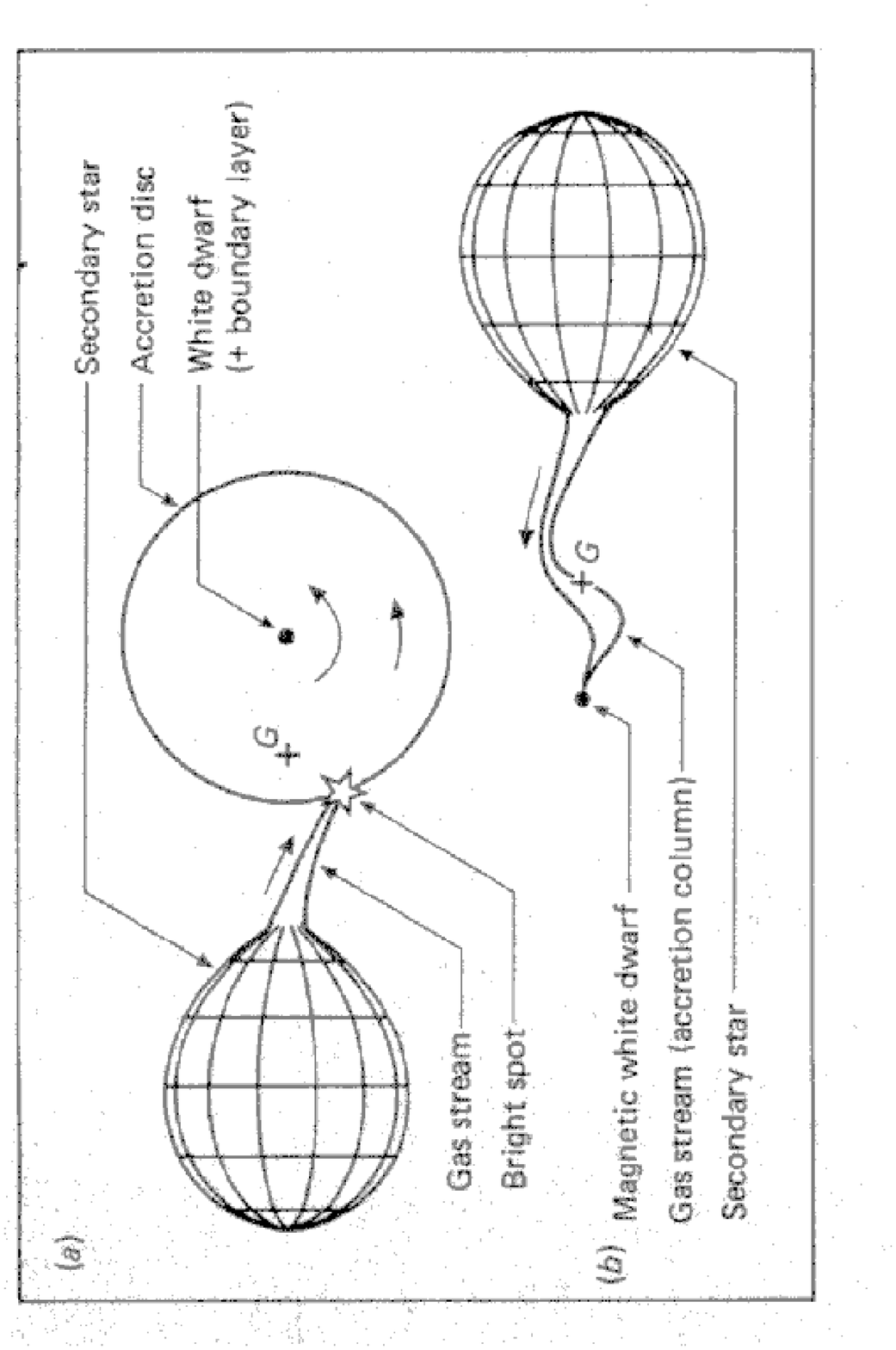}
\end{center}
\caption{\label{f_cv} The canonical model of a CV. A cool red secondary star and a white dwarf orbit around their common centre of mass, G; (a) in non-magnetic systems, there is an accretion disc which the gas stream hits at the bright spot; (b) in magnetic systems,  the gas stream is channelled along field lines onto the white dwarf. Reproduced from \cite{pringle85}, p. 132.}
\end{figure}

Variable stars have a great variety of behaviours on many different timescales, enabling astronomers to make key tests of the structure and evolution of stars. A subset of variable stars turns out to be in double star systems, and Cataclysmic Variables (CVs)\footnote{Excellent reviews giving considerably more detail than this article can be found in the books by Warner \cite{warner95} and Hellier \cite{hellier01}.} form a particularly interesting part of this subset, with orbital periods typically of less than half a day. The canonical model of these systems (figure~\ref{f_cv}) consists of a white dwarf star that is accreting material from a lower mass red dwarf star; the separation between the two stars is normally less than a few solar radii, so the interaction between the components is extreme and the red dwarf is tidally and rotationally distorted into a tear-drop shape known as a Roche lobe (Section~\ref{mdot}). If the white dwarf is non-magnetic, the accreting material spirals inwards onto the white dwarf surface through an accretion disc, landing preferentially in the white dwarf's equatorial zone; if the white dwarf is strongly magnetic, the accreting material (mainly ionized hydrogen) is forced to follow the field lines and crashes onto the surface near the magnetic poles of the white dwarf.

Amongst the many phenomena that are observed in these systems, the most dramatic are the outbursts that give the class its name. The largest outbursts are the {\it novae}: the star brightens by a factor of more than 10 million in a few days and then fades gradually over months to years. No ordinary nova has been observed to recur, although, as you will see, there are theoretical reasons for believing that they do, after an interval of 10$^4$ to 10$^5$ years. Less dramatic, but much more frequent, are the {\it dwarf nova} outbursts: here the star brightens by a factor of up to 100 in less than a day and fades over the next week or two -- however, they are all seen to recur, on timescales from a few weeks to a few decades. Some objects are known that share many of the characteristics of novae but have never shown an outburst; these are known as {\it nova-like} stars. Some of them may have been unrecognised novae in historical times. A third distinct class is formed by CVs whose white dwarf has a significant magnetic field -- these are now called {\it magnetic CVs} or MCVs; at least some of these have been observed to erupt as novae, but none of the strongly magnetic systems of the kind shown in figure \ref{f_cv}(b) has shown dwarf nova eruptions.

In this review, I shall give an overview of the properties of these objects, explaining what is currently known about them and how they are observed, with a particular emphasis on the recent development of imaging techniques for deducing their spatial structure from orbital variations in their appearance. I shall outline models for the outbursts and for the evolution of the systems with time, including a discussion of their probable origin, and I shall try to put the systems in context by indicating what can be learned from them about other stars and about particular aspects of physics.

\section{Some history}\label{history}

Novae have been known for many centuries, although it was only in the early decades of the 20th century that they were first clearly distinguished from the even more powerful {\it supernovae} (e.g. \cite{filippenko97}, \cite{hillebrandt00}), which I shall not discuss here. Dwarf novae were discovered much more recently, and quite gradually, with U Geminorum\footnote{Bright variable stars are denoted by one or two letters followed by the Latin name (in the genitive case: `U {\it of} [the constellation] Gemini') of the constellation in which they appear. The constellation name is usually abbreviated to 3 letters, but I shall spell out the constellation name in full at its first appearance.} being first seen in outburst in 1855 and the second one, SS Cygni, not being recognised until 1896. Their discovery as a distinct class from novae led to the designation of the latter as {\it classical novae}. Many dwarf novae were subsequently found in photographic surveys in the early twentieth century.

However, it was not until the 1960s that an initial suspicion was confirmed (e.g. \cite{kraft64}) that all novae and dwarf novae occur in binary stars. The evidence came partly from spectroscopic studies that revealed periodically varying radial velocities and partly from the development of photoelectric photometry that enabled the intensive study of the short-term flickering that is characteristic of many dwarf novae in quiescence. These studies fortuitously also discovered a number of eclipsing binaries.

Initial studies were concentrated in the optical part of the spectrum, which revealed mainly the disc and occasionally the white dwarf, but the development of red-sensitive CCDs in the 1970s and 1980s allowed the extension into the near infrared and the detection of an increasing number of the faint red secondary components (e.g. \cite{friend88}). Similarly the advent of ultraviolet satellites such as IUE allowed the more regular detection of the white dwarf components, culminating in the detailed studies made possible by the Hubble Space Telescope (e.g. \cite{hartley02}). Moving to even higher energies, X-ray satellites in the early 1970s sprang a surprise by finding a completely new population of X-ray binaries, some of which turned out to be CVs, mostly systems with magnetic white dwarfs.  The application of polarisation measurements allowed the direct detection of these magnetic fields in some CVs (e.g. \cite{tapia77}); the field strength can also be found from the presence of broad cyclotron lines in the visible and near infrared \cite{visvanathan79} and in some cases from Zeeman splitting. Nowadays, it is common for multi-wavelength studies to be employed to gain the maximum information about a system. One of the brightest CVs, AE Aquarii, has been observed over a very wide range of wavelength, from flares at GHz radio frequencies to pulsed $\gamma$-rays in the TeV energy range (e.g. \cite{warner95}, pp. 419--420).

X-ray observations have shown that there is an additional population of binaries that is closely related to CVs, but normally without a white dwarf. Instead, the accreting compact object is either a neutron star, with typical mass around 1.5\,M$_\odot$, or a black hole, with a mass typically in excess of around 5\,M$_\odot$. The more recently discovered very luminous supersoft X-ray sources do have a white dwarf as the accretor, but the accretion rate is some 100-1000 times higher than in CVs, leading to continuous nuclear burning on the white dwarf surface, and the mass-losing star is thought to be a sub-giant in most cases. These X-ray systems (see, e.g., \cite{lewin95}, \cite{greiner96}) share several of the characteristics of CVs, such as accretion discs, and many are included in the Ritter \&\ Kolb catalogue (\cite{ritter06}), but there is no space to discuss them here.

\section{Mass transfer}\label{mdot}

\begin{figure}[h]
\begin{center}
\includegraphics[width=8cm,angle=270]{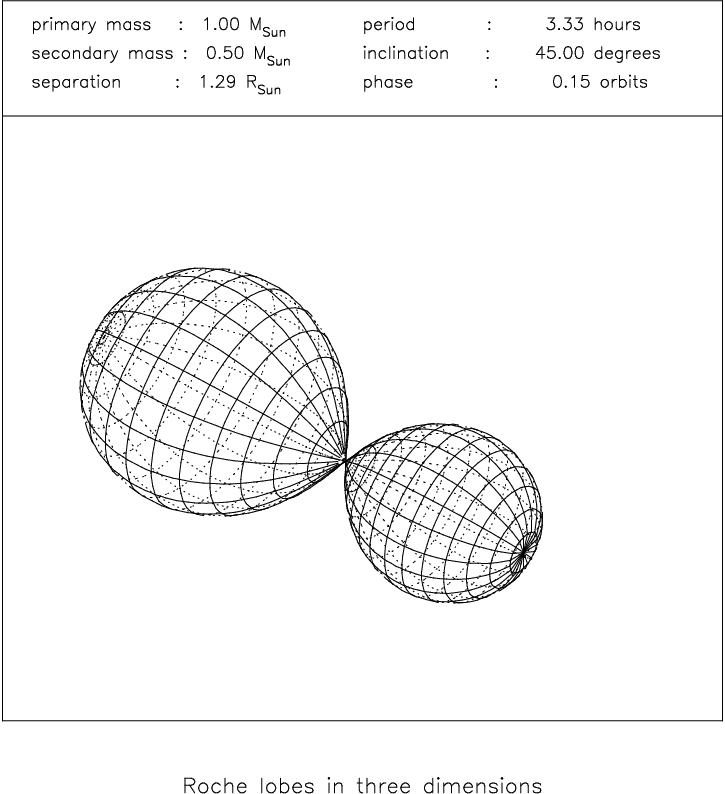}
\end{center}
\caption{\label{f_roche1} The Roche lobes of a binary system with mass ratio $q=0.5$: the more massive star occupies the larger lobe. If this were to represent a CV, the smaller lobe would be filled by the red dwarf, losing mass through the L$_1$ point where the two lobes touch, and the white dwarf would be at the centre of the larger lobe, which would also contain the accretion disc. Courtesy V.S. Dhillon, based on a program written by R.G.M. Rutten.}

\end{figure}

Most binary star systems consist of two well-separated components, with little or no interaction beyond their mutual gravitational attraction. However, in close systems tidal effects become important, leading to circularization of the orbits, and spinning up the stars' rotation speeds so that they are synchronous with the orbital motion. In a frame rotating with the orbital angular velocity, a test particle feels not only the gravitational attraction of the two stars but also a centrifugal force; the total force can be represented by a total potential. If the two stars are represented by point masses in a circular relative orbit, the potential is known as the {\it Roche potential}, and the potential surfaces approximately represent possible surfaces for the stars. There is a critical point on the line of centres between the two stars where the total force on a test particle is zero. This, and four other such points that I shall not discuss here, are known as the Lagrangian points, and the one between the stars is normally labelled L$_1$ and called the inner Lagrangian point. The potential surface through L$_1$ looks like a dumbbell (figure~\ref{f_roche1}) and the two tear-shaped lobes containing the two stars are known as the {\it Roche lobes}. In {\it detached binaries}, both stars are well within their Roche lobes. If one star fills its Roche lobe, the system is said to be {\it semi-detached}, while if both fill (or slightly over-fill) their lobes the term {\it contact binary} is used. Post-common-envelope binaries (Section~\ref{origin}) are generally detached, whereas CVs are semi-detached, with the red star filling its Roche lobe. (Contact binaries are also observed -- the so-called W Ursae Majoris stars -- but they are outside the scope of this article.)

The L$_1$ point is a saddle point of the potential and can be visualised as the lowest point on a mountain pass between the two valleys that represent the deep potential wells containing the two stars (the Roche lobes).  Material at the L$_1$ point is in equilibrium, but the slightest disturbance will send it into one or other of the Roche lobes. In semi-detached systems, when the Roche lobe of one star fills up there is a natural tendency for matter to flow over the saddle point into the Roche lobe of the other star. In CVs, the white dwarf is very small compared to its Roche lobe radius, so the accreting stream of matter from the Roche-lobe-filling red star, which has non-zero angular momentum relative to the white dwarf, does not fall directly onto its surface but goes into orbit around it. Collision of gas in successive orbits leads to dissipation and the formation of a disc; gas spirals inwards through the disc, and to conserve angular momentum a small amount of gas also moves outwards, carrying angular momentum to large radius, until the disc growth is limited by tidal forces from the red star. The maximum disc radius allowed by tidal interaction is about 90\%\ of the white dwarf's Roche lobe radius.

What effect does mass transfer have on the orbits of the stars? This depends on the mass ratio, as can be seen by a simple physical argument (more detailed discussions can be found in Pringle's chapter in \cite{pringle85} and in \cite{king88}). Consider a so-called {\it conservative} system, in which the total mass and angular momentum of the system are conserved. If mass is transferred from the more massive star to the less massive one, then the transferred mass is moving farther from the centre of mass of the system. This would have the effect of increasing the angular momentum of the system. However, if the system is isolated, the total angular momentum must be conserved, so the orbit must shrink to compensate; this has the effect of pushing the mass-losing star farther over its Roche lobe and increasing the mass transfer rate. It is clear that this process is unstable and in fact it leads to mass transfer on a dynamical timescale (see Section~\ref{origin} for an application). If, on the other hand, it is the less massive star that is losing mass, then the lost mass moves closer to the centre of mass of the system and the orbital radius must increase to conserve total angular momentum. This has the effect of cutting off mass transfer, so this kind of mass transfer will not persist (at least in a continuous fashion) unless either there is a steady increase in the star's radius (as would be the case for an evolving star) or the system is not conservative. Both kinds of behaviour are seen in CVs, as discussed in Sections \ref{aml} and \ref{pdist}.

It is worth noting here that a more detailed discussion (e.g. \cite{politano96}) has to take into account also the response of the mass-losing star to the mass loss: does its radius increase or decrease? on what timescale? and how does the rate of radius change compare to the rate at which the Roche lobe radius is changing? The great majority of CVs have mass ratios that fall in the stable part of parameter space; those few that appear to lie in the unstable region have rather uncertain orbital parameters \cite{smith06}, and it is believed that all CVs are stable to mass transfer from the secondary star.

\section{The CV zoo}\label{zoo}

The three main types of CV mentioned in the Introduction each have various sub-classes, which I will introduce here, together with a related group of stars that are similar to CVs but not one of the three main types.

\subsection{Novae}\label{cn}

Novae are all fairly similar, differing mainly in the rate of rise and decay. The `speed class' is described in terms of the parameter $t_3$, which is the time taken for the light to drop by 3 magnitudes\footnote{The magnitude, $m$, of a star is a logarithmic measure of its brightness, defined by $m = -2.5\log_{10}F + \mbox{constant}$ where $F$ is the measured flux density in the particular waveband being considered. An increase of brightness by a factor of 100 corresponds to a decrease of 5 magnitudes. The constant is chosen so that the scale corresponds roughly to a scale originally introduced by Hipparchus some 2,000 years ago; the faintest stars visible to the naked eye have a magnitude of around +6.} from maximum light. For very fast novae this can be as short as a few days, but for some very slow novae it can be many months. Faster novae also rise faster, and have much larger amplitudes of outburst. Novae expel large shells of gas that expand at speeds of 1000s of km\,s$^{-1}$ and gradually fade into the general interstellar medium.

The classical novae (CNs) have not been seen to rise more than once, but there is a small class of {\it recurrent novae} (RN) that seem to recur on timescales of decades or centuries. Fewer than a dozen are known, and around half of these (the T Coronae Borealis subclass) have a red giant as the secondary star, so they do not strictly fall into the CV class as I have defined it. T Pyxidis is a unique RN that has an orbital period (2.38 hr) characteristic of CVs but a slow decline; the nature of its secondary is unknown, but it appears to have a very  massive white dwarf and it has been suggested (e.g. \cite{selvelli03}) as an example of a progenitor for a Type Ia supernova explosion. Members of the U Scorpii subclass of RN also have relatively short orbital periods ($\sim$1 day), He-dominated discs in quiescence and probably evolved secondaries; their rise and fall during eruption are amongst the fastest known.

The post-nova settles down as a CV with a fairly high mass-transfer rate. Systems that look similar in appearance to post-novae, but have not been recorded in outburst, are usually called nova-like variables (NLs): they show low-level variability but no outbursts and are probably old novae whose outbursts occurred before records began, or went unrecorded for some other reason.

\subsection{Dwarf novae}\label{dn}

\begin{figure}
\begin{center}
\includegraphics[width=15.8cm,angle=270]{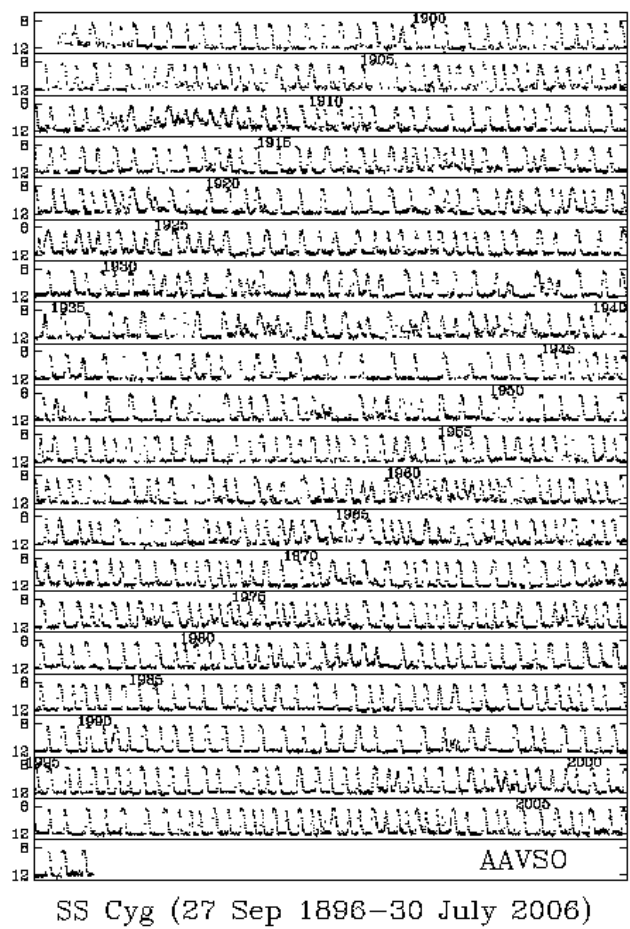}
\end{center}
\caption{\label{f_sscyg}The light curve of SS Cygni from 1896 to
2006, based on amateur data from the American Association of Variable Star
Observers (AAVSO). The ordinate scale is visual magnitude. Courtesy J. Cannizzo.}
\end{figure}

The canonical dwarf novae are the U Geminorum, or U Gem, stars, named after the prototype, first observed in outburst as early as 1855 (\cite{warner95}, p.3). These show semi-regular outbursts with a typical timescale between them that ranges from weeks to years and a typical amplitude of 4-5 magnitudes (a factor of 40-100 in brightness). Many dwarf novae have been extensively observed by amateurs, and the best observed U Gem star is SS Cygni, which has been well observed for over a century. Its light curve is shown in figure~\ref{f_sscyg}. A first impression of regularity soon gives way to the realisation that there is a great variety of shapes and durations of outbursts. The same is true for all dwarf novae, and this wide variation is a key challenge for theorists.

\begin{figure}
\begin{center}
\includegraphics[width=8cm,angle=270]{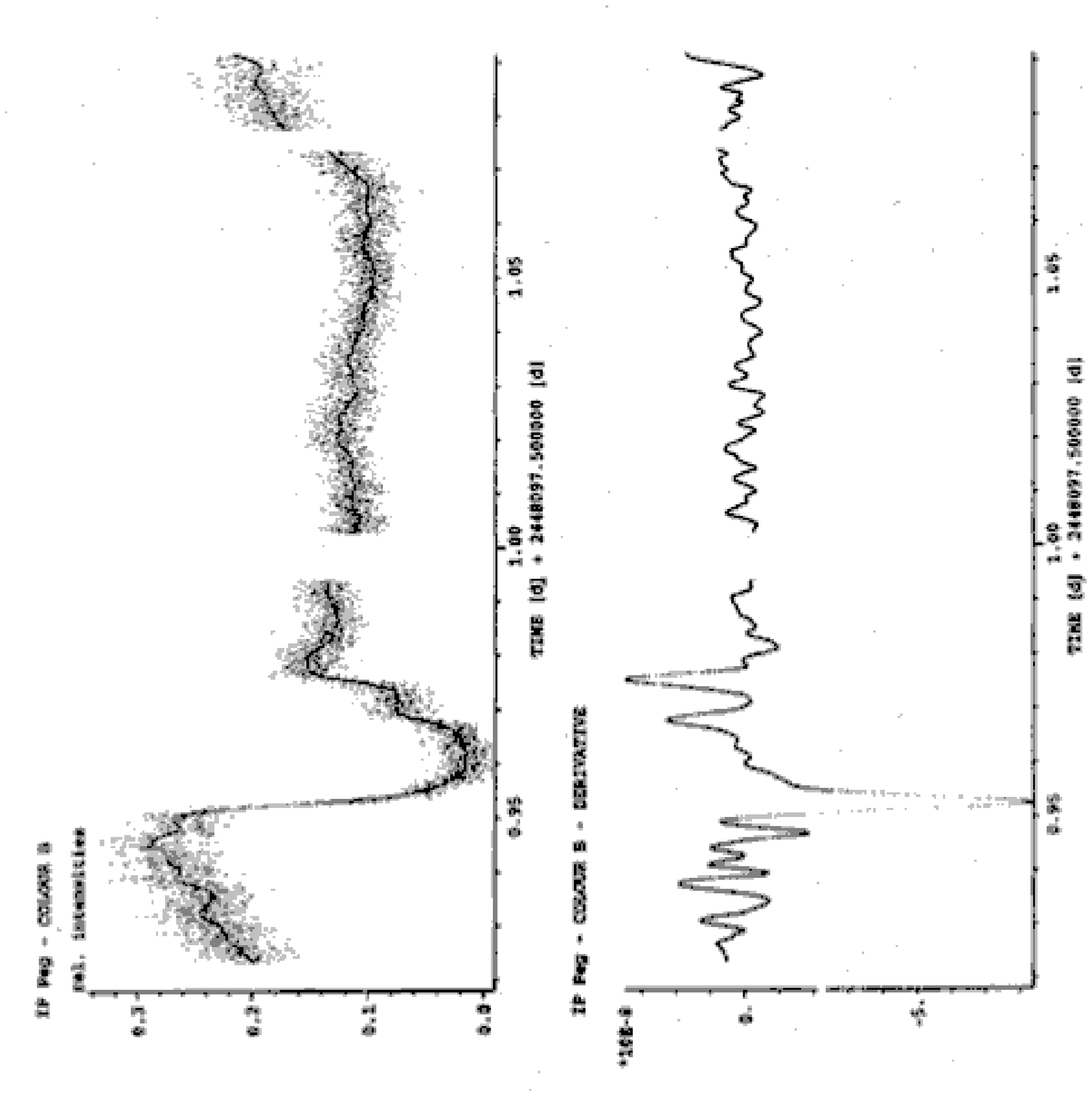}
\end{center}
\caption{\label{f_hump}The $B$ magnitude light curve of an eclipse of IP Pegasi, with a spline fit (\cite{wolf93}); the vertical axis is relative intensity and the horizontal axis is time in days. The lower panel shows the derivative of the spline fit and the sharp peaks correspond to the timings of the eclipses of the bright spot and the white dwarf. Note the clear pre-eclipse hump, and the larger scatter (`flickering') out of eclipse.}
\end{figure}

The light curves also show variability on an orbital timescale, from low-amplitude flickering on timescales of seconds or minutes through smooth variations arising from the varying visible surface area of the tidally distorted secondary star to eclipses in some systems, and pre-eclipse humps that arise from the bright spot where the accretion stream runs into the disc (figures \ref{f_cv} and \ref{f_hump}).

There are two important sub-classes of dwarf novae. The Z Camelopardalis, or Z Cam, stars (first discovered in 1904) alternate between normal outbursts and periods of inaction, known as `standstills',  where the star's brightness remains static at a value about one third of the way down from maximum brightness. These periods of constant light may last for many months, but their length varies considerably, even for a given star. While in standstill, Z Cam stars resemble the nova-like systems that never show outbursts, but they are somewhat less luminous.

The SU Ursae Majoris, or SU UMa, stars show mostly normal outbursts, but every now and again they produce a superoutburst, which lasts significantly longer and has a slightly larger amplitude. These superoutbursts actually occur rather more regularly than the normal outbursts, and the orbital light curve during superoutburst is characterised by superhumps -- like the humps seen in normal dwarf novae, except that superhumps drift in orbital phase; these are believed to be the signal of an elliptical disc whose long axis is precessing around the orbit (Section \ref{suuma_ob}).

\subsection{Magnetic CVs}\label{mcv}

The magnetic white dwarfs in CVs were first detected by polarisation measurements. In 1976, the known NL system AM Herculis was both found to be the counterpart of an X-ray source and discovered to be polarised at optical wavelengths (see \cite{warner95}, p.307, for more detail). The key characteristic deduced from the polarisation is a very strong magnetic field, with magnitude typically in the range 1000--8000\,T (10--80\,MG; an extreme example, AR UMa, has a 230\,MG field). In some systems, the field strength can be measured by Zeeman splitting of absorption lines from the white dwarf, and in others the spectrum shows a series of humps in the optical or near infra-red, corresponding to harmonics of cyclotron emission from electrons spiralling down magnetic field lines onto the white dwarf surface; although these humps are often hard to detect, and may only appear at some orbital phases, they do provide a third way of measuring the magnetic field strength.

These systems are commonly known as {\it polars}, or AM Her stars, and the field is strong enough to force the white dwarf to spin with the orbital period; this results from interaction with the (much weaker) magnetic field of the secondary (see \cite{campbell97} for a full discussion). In the polars, there is no accretion disc -- the magnetic field forms a magnetosphere around the white dwarf and when the ionized accretion stream encounters the magnetosphere the flow becomes tied to the magnetic field lines and material is directed along the converging field lines, accelerating as it falls and forming a shock at the top of an accretion column near the magnetic poles of the white dwarf. Enough energy is released on impact to make the systems X-ray sources.

Other magnetic systems have weaker magnetic fields, so the magnetosphere is smaller and a disc may form outside it if the field is weak enough. Material then meets the magnetosphere at all points on the inner edge of the disc, and the flow becomes an {\it accretion curtain} rather than a single converging stream. Another consequence of the weaker field is that synchronous rotation is no longer enforced, and the white dwarf is in general spinning about ten times faster than the orbital frequency, with periods measured in minutes rather than hours; this faster spin results from the accretion of high angular momentum material from the secondary star. Because they fall between polars and non-magnetic systems, these systems are generally called {\it intermediate polars} (IPs for short), or sometimes DQ Her stars, although the type star DQ Her, the first IP to be discovered, is actually atypical of the class, with the very short white dwarf spin period of 71\,s. Some authors (e.g. Warner \cite{warner95}) regard DQ Her systems as a subclass of IPs, with particularly rapidly spinning white dwarfs and no detectable hard X-ray emission. However, as more IPs are discovered, with a range of spin periods, it may be more appropriate to regard DQ Her as being at one extreme of a continuous distribution of spin speeds. Because of the presence of a disc (see Section \ref{dn_ob}), some IPs show dwarf nova type outbursts.

\subsection{AM CVn stars}\label{ac}

Although not strictly CVs as defined earlier, the rare AM Canum Venaticorum (AM CVn) stars are sufficiently closely related that they are often included in catalogues and discussions. They have extremely short orbital periods, between 10 and 65 minutes, and their spectra show no evidence for hydrogen. They appear to be helium-rich versions of CVs, and they will be discussed further at the end of Section~\ref{pdist}.

\section{Observing CVs}

\subsection{Photometry}\label{photom}

For nearly a century, in addition to the photographic studies by professionals that normally led to the discovery of new CVs, visual observations by amateurs played a key role in their long-term monitoring, and most attention was focussed on the obvious outbursts. A turning point came with the introduction of photoelectric photometry and in particular the development of the 1P21 photometer that was to become the workhorse of professional variable star observing from the 1940s until CCDs began to take over in the late 1970s. This new method of observing, especially when high-speed pulse-counting photometers became available \cite{warner88}, enabled more intensive monitoring of CVs on shorter timescales and revealed many variations on an orbital timescale. It rapidly became apparent that one of the defining characteristics of CVs was the presence of low-amplitude (0.01 to 0.2 mag) flickering on timescales from 10s of seconds to a few minutes. Detailed examination of eclipse profiles showed that these also varied, this time from one orbit to the next, and that there was often a pre-eclipse hump in the light curve (cf. Section~\ref{dn}).

Flickering turned out to provide important clues to what is going on in CVs. It was noticed that the amplitude of the flickering was a maximum at the peak of the pre-eclipse hump, suggesting that the flickering arose mainly from the bright spot where the accretion stream struck the disc, and that flickering vanished altogether during eclipse. This both confirmed that the bright spot was being eclipsed and, more importantly although less directly, implied that it was the accretion disc itself that was brightening during an outburst. Before that realisation, it had generally been thought that the seat of the dwarf nova outburst was one of the component stars.

Another key result of photoelectric photometry was the discovery by Merle Walker \cite{walker56} of a precisely periodic variation with period 71.1s in DQ Herculis; as discussed in Section~\ref{mcv}, this is now interpreted as the spin period of the white dwarf, but it was some 20 years before the next precisely periodic signal was found. In the meantime, a rather less stable variability had been discovered, now known as dwarf nova oscillations, or DNOs. These have a period stability parameter $Q=|\dot{P}|^{-1}$ in the range $10^4-10^6$, as opposed to $Q>10^{12}$ for the white dwarf rotation in DQ Her stars; unlike flickering, DNOs occur only during outburst, and have such low amplitude that they were initially detectable only via power spectrum analysis. The origin of DNOs is still obscure, although most models involve magnetic accretion. A similar but distinct phenomenon, with rather larger amplitude but much lower $Q\ (\simeq1-10)$ and generally longer periods by a factor of a few, is that known as quasi-periodic oscillations, or QPOs. The mechanism is again uncertain, but may be related to oscillations or travelling waves in the accretion disc. Similar QPOs, but with much higher frequencies, occur in binary systems containing neutron stars and black holes; there is a clear correlation between the high and low frequency QPOs that occur in such systems, with the high frequency being roughly 15 times the low frequency. Remarkably, the DNO oscillations in CVs fit onto the same correlation, with $P_{\rm QPO}/P_{\rm DNO} = 15$ (see Figure 30 in \cite{warner03}), suggesting that a similar mechanism may be operating in all these systems.

\subsection{Spectroscopy}\label{spectra}

One of the clues to the binary nature of old novae and dwarf novae was the discovery of composite spectra: there appeared to be several components present, of different temperature. In the canonical CV, the dominant contributor to the light in the optical range is emission from the accretion disc, consisting typically either of optically thin continuum emission and strong emission lines, mainly of hydrogen and helium, or of optically thick continuum emission plus the same lines in absorption. The continuum emission was originally modelled as black-body emission from a series of rings, each of a constant temperature, with the temperature decreasing smoothly outwards from the centre of the disc. More recent models represent the spectrum of each ring by a stellar atmosphere model of the appropriate surface gravity and effective temperature. In addition, detailed models of the emission line strengths have shown (e.g. \cite{robinson93}) that in most cases emission lines of the observed strengths cannot arise from a disc that is optically thin in the continuum; it is thought that emission lines arise either from the upper, chromospheric layers of the disc, where there is a temperature inversion (temperature reaches a minimum and then increases outwards), or by irradiation from the white dwarf or inner disc, causing excitation/ionization; I shall return to this point at the end of Section \ref{suuma_ob}.

Towards the red end of the optical range, it is often possible to detect the red companion, whose continuum emission increases strongly towards the infrared, where the companion's spectrum often dominates. The spectral type of the red secondary may be as early as G0, showing many absorption lines of iron, or as late as M8 or even into the extremely cool stars now categorised as L or T types (\cite{kirkpatrick05}). The coolest stars show strong molecular absorption bands; in the M stars, these comprise TiO in the optical and H$_2$O and CO in the infrared. The strong NaI absorption doublet near 820\,nm in late K and M stars has been exploited by many people to detect the secondary and to determine its orbital motion (e.g. \cite{friend88}, \cite{friend90}).

In a few systems the optical spectrum of the white dwarf may also be detected, with absorption lines of hydrogen, which are strongly pressure broadened because of the high surface gravity. The white dwarf spectrum often becomes dominant in the ultraviolet, where it has been well studied using the IUE, HST and FUSE spacecraft. The white dwarf temperature generally lies in the range 10,000 to 40,000\,K.

\subsection{Astrotomography}\label{tomography}

\begin{figure}[h]
\begin{center}
\includegraphics[width=10cm,angle=270]{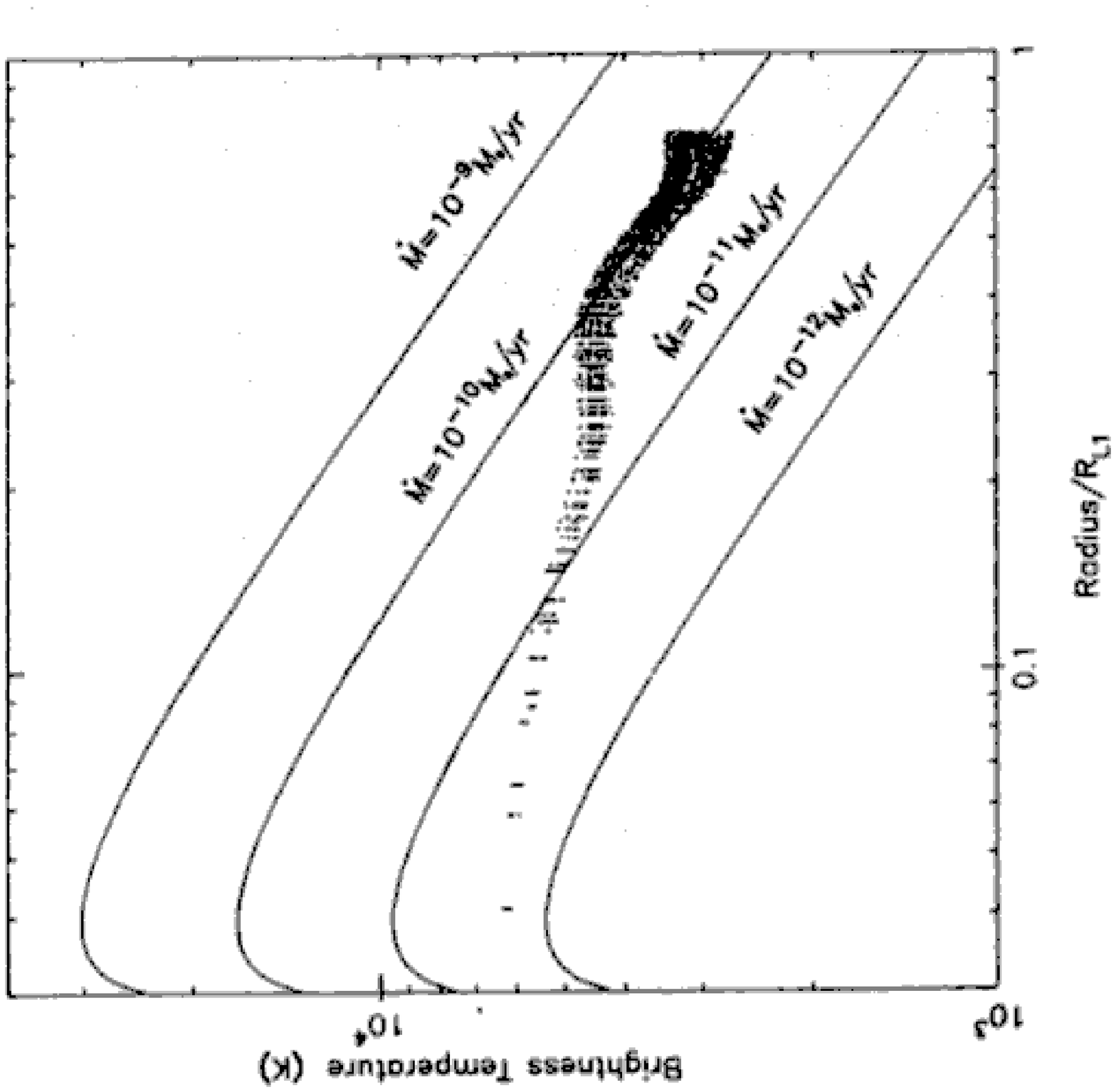}
\end{center}
\caption{\label{f_zcha}The temperature profile of the disc in Z Cha in quiescence, compared to profiles predicted by steady-state disc models with various mass-transfer rates (reproduced from \cite{wood86}, with permission).}
\end{figure}

One of the most exciting developments of the last 20 years has been the mapping of emission patterns in the accretion flow and more recently on the secondary star. Of course, this cannot be done by direct imaging: at a distance of 1\,kpc, the typical separation of the stars of around 10$^9$\,m corresponds to an angle of less than 10 microarcsec, and it is desirable to see features on scales at least 100 times smaller. The imaging must therefore be done indirectly, using the varying geometry as the stars orbit around one another: time variations in emission intensity can be interpreted as spatial variations in the disc or on the secondary star. Perhaps the simplest example is what is termed {\it eclipse mapping}. In eclipsing CVs, the secondary star passes in front of some or all of the accretion disc, causing the disc light to decrease as it is progressively obscured. The shape of the eclipse profile is determined by the brightness variations across the disc, and in principle these brightness variations can be reconstructed from the eclipse profile. However, the eclipse profile is one-dimensional while the disc brightness pattern is two-dimensional, so the reconstruction process is not unique without an additional constraint. The procedure devised by Horne \cite{horne85a} was to make use of the concept of image entropy (\cite{gull78}, \cite{bryan80}) and to maximise the entropy subject to some constraint; the two constraints normally chosen are to make the intensity distribution maximally uniform or maximally axisymmetric -- the latter is most often used because it allows the radial dependence of the intensity to be reconstructed. Horne \&\ Cook \cite{horne85b} found that the temperature profile in Z Chamaeleontis in outburst matched quite well the prediction from a simple steady-state disc. Later observations of Z Cha in quiescence \cite{wood86} showed a much flatter temperature profile (figure~\ref{f_zcha}); this is explained by the inability of an optically thin disc to reach a steady state.

By extending this technique from photometry to spectroscopy, and looking at how the spectra vary through an eclipse, it has also proved possible to deduce the changes in the spectrum of the disc as a function of radius, as explained in detail by Baptista \cite{baptista01}. As expected, these changes are generally consistent with a decrease in the effective temperature outwards, with the continuum emission becoming progressively redder and fainter towards larger radii. Systems with high mass transfer show optically thick spectra near the disc centre, with absorption features, but these make a transition to optically thin spectra, with emission lines, at larger radii. However, there were also unexpected features revealed by the additional spectral information: at least one system shows evidence in the line profiles of gas outflow from the disc, while other systems reveal that the accretion stream, whose spectrum is different from that of the disc, skims over the surface of the disc after overflowing the disc edge at the bright spot where the stream first impacts the disc. A step beyond spectroscopic eclipse mapping is now being taken by direct mapping of physical parameters, such as temperature and surface density, instead of intensity \cite{vrielmann01}.

\begin{figure}
\begin{center}
\includegraphics[width=12cm,angle=180]{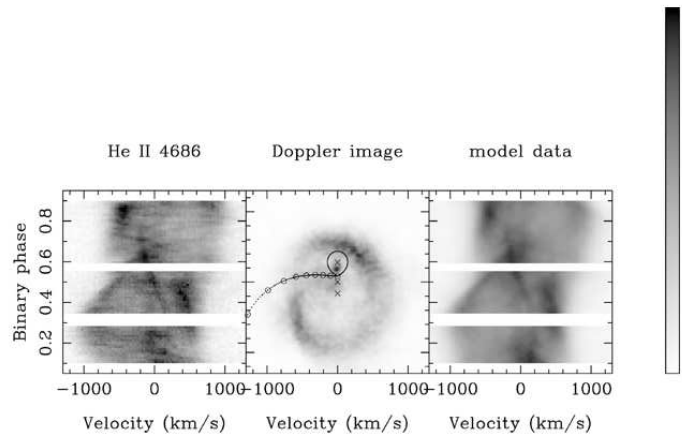}
\end{center}
\caption{\label{f_doppler}Doppler map of IP Pegasi in the ionized helium (HeII) emission line at 468.6\,nm. The left-hand panel shows a set of spectra at most phases around the binary orbit -- this presentation, with wavelength (in velocity units) on the x-axis and binary phase on the y-axis, is known as a trailed spectrogram. White denotes low intensity and black high intensity -- so the black parts of the spectrogram denote the emission line. The central panel shows the resultant Doppler image, with crosses marking the positions of the white dwarf, the centre of mass of the system and the red dwarf, whose outline is also shown. The curve marks the position of the accretion stream in velocity space, assuming it to be in free fall. Emission is visible from the irradiated face of the secondary; there is also non-axisymmetric emission from the disc, which is interpreted as showing the presence of spiral arms. The third panel shows the trailed spectrogram that would result from that interpretation; it matches the observed data well. (From \cite{harlaftis99}, with permission.)}

\end{figure}

Eclipse mapping is, of course, only applicable to the relatively small fraction of systems whose orbits are viewed almost edge-on. However, it was soon realised that variations in the emission line profiles around the orbit could be used to map the source of the emission. The observed emission lines from the disc are usually broad, and often double-peaked. The width of the profiles arises from the large range ($\sim10^3$\,km\,s$^{-1}$) of orbital velocities in the disc, which to a first approximation is a set of particles in Keplerian orbits around the white dwarf. The profiles thus contain information not only about the intensity of the emission but also about the velocity field in the disc. Horne and Marsh (\cite{horne86}, \cite{marsh88}) developed the technique of {\it Doppler tomography} to invert the line profiles and deduce information about the intensity and the velocity field. A maximum entropy method is again used, but this time to smooth over incomplete data, noise and instrumental blurring. Unlike eclipse mapping, the result is not sensitive to constraints of smoothness or axial symmetry, essentially because there is now information from the whole orbit instead of just from the time of eclipse. If it is assumed that the disc is axisymmetric and in purely Keplerian rotation, then the velocity field can be translated into position in the disc and a spatial map can be deduced. However, the assumption of pure Keplerian motion is unrealistically restrictive, and some discs are certainly non-axisymmetric, so it is usual instead to express the maps as intensity at points in velocity space, in the frame rotating with the orbital frequency. These `Doppler maps' are therefore not so straightforward to interpret as a spatial map would be; for example, the disc is turned `inside out', because the low velocity points are far out in the disc and the high velocity points correspond to the rapidly spinning inner disc. To aid interpretation, it is usual to overlay on the maps the positions in velocity space of the two stellar components (including the extended shape of the secondary, assumed to be uniformly rotating at the orbital frequency) and a line showing the path of the accretion stream in velocity space. An example of a Doppler map is shown in figure~\ref{f_doppler}. One of the most intriguing phenomena revealed by Doppler mapping has been the existence of spiral patterns in the discs of a small number of CVs (\cite{steeghs01}, \cite{hartley05}), both dwarf novae in outburst and novalikes (which are effectively in continuous outburst); see the end of Section \ref{suuma_ob} for further discussion.

Doppler maps have been used to show that some emission comes from the secondary star; the corresponding line profiles normally show a narrow component, corresponding to the fact that the secondary star has a much smaller range of velocities associated with it -- just the broadening arising from the star's rotation speed of around 100\,km\,s$^{-1}$. However, Doppler tomography does not have the resolution to map the surface of the secondary, and in any case the spectrum of the secondary mostly contains absorption lines rather than the emission lines used in Doppler tomography. Nonetheless, it is possible to use tomographic techniques to map the surface of the secondary, following methods developed for single, rapidly rotating stars by Collier Cameron, Donati and their collaborators (see \cite{cameron01}, \cite{donati01} for reviews). The additional complication for stars in binary systems is that they are not axisymmetric, because of the tidal distortion from the companion. It is necessary to take into account the full Roche geometry of the secondary star's surface, so this technique is called {\it Roche tomography}.

Before I describe that, let me note the problem that first led people to try to work out the pattern of intensity over the surface of the secondary. If the secondary were uniformly bright, then the centre of light would coincide with the centre of mass of the star and radial velocity measurements would correctly represent the orbital motion of the star. However, it was realised by many people that irradiation from the disc and white dwarf would heat the inner face of the secondary, and might well ionize the neutral atoms whose lines were being used to find the radial velocity. This would mean a lack of contribution to the line profile from that part of the star, and a displacement of the centre of light towards the unirradiated side of the star. Because that side of the star would be rotating towards the observer when the star was approaching the observer, this would lead to an artificially larger amplitude for the radial velocity curve. Similarly, radial velocity curves taken from the narrow emission lines arising from the irradiated face would have too small an amplitude -- and in very few systems can both be observed. Some other way was needed to correct for the effects of irradiation, and this led to simple models that could be used to correct for irradiation and in some cases to provide crude maps of the surface brightness (e.g. \cite{davey92}, \cite{davey96}). By the time of the latter paper, Rutten and Dhillon (\cite{rutten94}) had developed the first code to take the full geometry into account, and an early Roche tomogram of the polar AM Her, using their code, appears in \cite{davey96}.

The current version of the code is now good enough to reveal starspots rather than just the irradiation pattern. It depends on the fact that an absorption line profile is made up of contributions from the entire visible hemisphere of the star, with different velocities and intensities. If the hemisphere were uniformly bright, the profile would represent the rotational broadening, plus some distortion arising from the tidally-distorted surface shape. The presence of a localised dark spot or region on the surface, occurring therefore at a particular radial velocity, introduces a bump in the line profile at the wavelength corresponding to that velocity (see Fig. 1 in \cite{cameron01}). As the star moves in its orbit, the spot appears to move across the surface, just as sunspots do as the Sun rotates, and the corresponding velocity and wavelength of the bump in the profile change: the bump appears to move across the profile. The orbital phases at which the bump first appears and then disappears allow a determination of the longitude of the spot. If the star were observed in the plane of the orbit, a spot at the equator would show an amplitude of motion corresponding to the full rotation speed of the star; a star at the pole would show no motion at all. This makes it possible in principle also to determine the latitude of the spot, from the amplitude of its motion and a knowledge of the orbital inclination. In practice, a single line profile is often too noisy for the bump even to be visible, never mind measured, and it has only been possible to determine the general pattern of irradiation with a single line analysis (\cite{dhillon01}).

\begin{figure}[h]
\begin{center}
\includegraphics[width=12cm,angle=180]{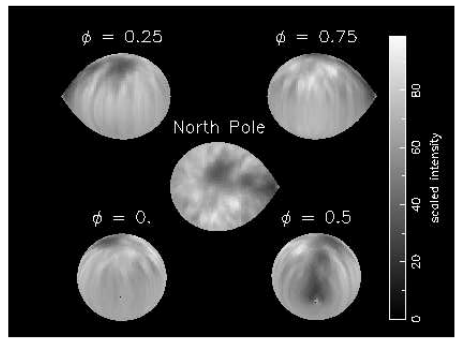}
\end{center}
\caption{\label{f_aeaqr} Roche tomogram of AE Aqr (reproduced from \cite{watson06}, with permission). Dark grey areas represent regions of low absorption, and are either star-spots or irradiated regions. The system is shown at four different orbital phases, as seen by an observer at the orbital inclination of 66$^\circ$. The central map shows the system as seen looking down on the north pole.}

\end{figure}

The breakthrough was the development in the single star field of the procedure of {\it least squares deconvolution} (LSD), which essentially takes the mean profile of many hundreds of absorption lines. This in principle increases the S/N ratio in the profile by the square root of the number of lines; in practice, the increase is about half that, but still very significant -- a factor of 15 or more if 900 lines are used. Once the mean profile has been produced as a function of orbital phase, the inversion to obtain a map proceeds, as in other tomographic methods, by defining an image entropy relative to a default map and looking for the maximum entropy solution. The default map used in current applications is a moving uniform map, where at each stage in the calculation the uniform value is the average value in the solution map. Even with the refinement of LSD, only the brightest CVs are accessible with 4-m class telescopes; the first successful detection of a starspot was on AE Aqr, which clearly has a spot near the north pole and may have smaller spots nearer the equator (figure~\ref{f_aeaqr}; see also \cite{watson06}). Unpublished data for BV Centauri (Watson, private communication) and RU Pegasi (Dunford, private communication) suggest that these systems also have polar spots, and that in all three systems the spots are displaced towards the trailing hemisphere, possibly as a result of Coriolis forces (Watson, private communication). Polar spots are also common features on rapidly rotating single stars, although not seen on the slowly rotating Sun, and their presence on the even more rapidly rotating CV secondaries suggests that the magnetic field structure changes with rotation, with spots being able to form over an increasing range of latitude as the rotation rate increases.

All the above techniques are now given the generic title of {\it astrotomography} and a workshop on the subject in 2000 gave rise to a very useful review volume \cite{boffinetal01} that covers a wider range of applications than there is space to discuss here.

\subsection{Observational surveys}

Astrotomography looks in great detail at a very small number of systems. At the other end of the spectrum, there is increasing statistical information about the CV population, as a result of observations over a wider spectral range and of new large-scale surveys that are providing for the first time large homogeneous samples of CVs with carefully defined selection criteria. One of the difficulties in comparing the observed population with population synthesis models (Section \ref{popsyn}) has always been that the observational selection biases are poorly known; soon it should be possible to model the biases in these new samples. The histogram of the period distribution shown in figure \ref{f_period} is based on about 600 systems, with around half of the currently known magnetic systems (and a good many other CVs) being discovered by the ROSAT X-ray satellite. Recent important surveys are the 2dF Quasar Survey, which has yielded more than 20 new CVs, and the Sloan Digital Sky Survey, which is expected to yield some 200 new CVs by the time it is completed. A useful recent summary is given by G\"{a}nsicke \cite{gaensicke05}.

\section{Outburst mechanisms}

\subsection{Classical Novae}\label{cn_ob}

Nova outbursts are essentially giant thermonuclear explosions. The white dwarf has lost the great majority of its original hydrogen as a result of its previous evolution (cf. Section~\ref{origin}) and it is either mostly helium, if its mass is less than about 0.5\,M$_\odot$, or mostly a mixture of carbon and oxygen, if its mass is between about 0.5 and 1.2\,M$_\odot$. White dwarfs with masses greater than 1\,M$_\odot$ may have an oxygen/neon core. However, the material that it is accreting from its companion is largely hydrogen, so a layer of fresh hydrogen gradually builds up on the surface. As the density of the layer increases under the weight of the accreting material, the electrons in the ionized hydrogen become degenerate, and the pressure of the degenerate electrons becomes the dominant pressure in the layer, except very near the surface. As the layer thickens, the temperature at the base of the layer increases, until eventually it becomes high enough for hydrogen fusion to occur. In a normal stellar core, where thermal gas pressure dominates, the onset of hydrogen burning occurs gently, because the resultant rise in temperature causes a corresponding rise in pressure that expands and cools the burning layer, moderating the burning rate. However, in a degenerate gas the pressure is independent of temperature, so the onset of burning causes a rise in temperature that is not moderated by thermal expansion. Because the burning rate is very sensitive to temperature ($\propto T^{40}$), this leads to a thermal runaway and hence to an explosive ignition of hydrogen burning. The violence of the explosion ejects the entire layer of hydrogen in a nova shell, leaving the white dwarf once again essentially hydrogen-free. The whole process starts again, and the explosion will repeat on a timescale that theoretical calculations suggest is about 10,000 to 100,000 years (\cite{hernanz05}).

Current calculations yield a mass for the ejected shell that is comparable to the masses of observed nova shells. The difficulty with the mechanism is getting the mass transfer rate right: if the mass transfer rate is too high, it will cause compressional heating of the accreting layer, which will inhibit the onset of degeneracy and may prevent envelope ejection altogether. An upper limit of around 10$^{-8}$\,M$_\odot$\,y$^{-1}$ is required (e.g. \cite{livio92}), which is fortunately comparable to the largest values deduced from NL systems for pre-novae. Note that the inhibition of envelope ejection may lead to the white dwarf  mass increasing sufficiently to take it over the critical Chandrasekhar mass of 1.4\,M$_\odot$; this could trigger a Type Ia supernova explosion (cf. \cite{hillebrandt00}). However, in normal novae, models suggest that the mass of the white dwarf if anything decreases as a result of successive nova explosions, so it is currently believed that novae will not become supernovae.

\subsection{Dwarf Novae}\label{dn_ob}

The outbursts in dwarf novae are both simpler and more complicated than those in novae. They are simpler because the energy source for the explosion is essentially just the gravitational field of the white dwarf -- but the details of the instability that leads to the outburst are complex. In essence, what happens is that mass transferred from the companion is stored for a while in the disc until the density in the disc reaches a critical value. The stored mass is then dumped rapidly onto the white dwarf, releasing a great deal of energy as it falls down the deep gravitational potential well. The original proposal for this model \cite{osaki74} said little more than that -- and it was seven years before it was agreed what actually triggered the outburst at a particular density.

To explain the trigger, I need first to look briefly at the properties of accretion discs (for extensive reviews, see \cite{pringle81}, \cite{papaloizou95}, \cite{lin96}). The key property is viscosity: without it, particles in the disc would simply orbit the white dwarf indefinitely and there would be no accretion. It is viscosity, and the associated viscous torques between annuli in the disc, that allows angular momentum to be transferred outwards and mass to spiral inwards. Any original ring of particles will thus spread out into a disc, with the outer radius being determined by tidal torques from the companion and with accretion occurring from a boundary layer very close to the white dwarf surface. However, it was clear from the start that ordinary molecular viscosity is completely inadequate to account for the observed properties of discs and some kind of turbulent viscosity is invoked. The characteristic scale of the turbulence must be less than the disc thickness, $H$, and the characteristic turbulent speed is expected to be less than the speed of sound, $c_s$, since there is no strong evidence for turbulent shocks. The viscosity, $\nu$, is therefore often parametrised by writing it as $\nu=\alpha c_s H$, placing all the uncertainties in the unknown parameter $\alpha$, taken to be less than 1. This is the standard alpha-disc model of accretion discs, used by most authors, despite the fact that there is still no reliable way of calculating $\alpha$. The general belief now is that angular momentum transport in discs actually results from a magneto-rotational instability, as proposed by Balbus and Hawley (\cite{balbus91}, \cite{balbus98}; see also Section \ref{conc}). This mimics viscosity, but does not strictly speaking generate turbulence; however, an effective $\alpha$ can be estimated and is sometimes used to compare the detailed numerical results with those from alpha-disc theory.

Because of viscosity, and the resultant diffusion of material, the disc acts as a mass transfer channel between the mass-losing star and the white dwarf. However, the rate of mass flow through the disc, $F_M$, is set by the value of the viscosity, and will in general not be equal to the rate of mass transfer from the red dwarf, $\dot{M}$. If $F_M<\dot{M}$, then mass will build up in the disc, whereas if $F_M>\dot{M}$ mass will drain out of the disc. If the viscosity were a monotonically decreasing function of the density in the disc, then increasing the mass would decrease the viscosity and hence decrease $F_M$, increasing any imbalance; in that case, no equilibrium would exist. On the other hand, if the viscosity were a monotonically increasing function of the density in the disc, then increasing the mass would also increase the viscosity, and hence the mass flow rate, and an equilibrium state with $F_M=\dot{M}$ could be reached. Such an equilibrium would be stable (on both thermal and viscous timescales; cf. \cite{pringle81}, \cite{frank02}) and there would then be no outbursts. The insight that resolved this paradox (\cite{bath82}, first described in a talk by Pringle in 1981) was that a viscosity-density relation which was double-valued for at least a range of density could give rise to a limit cycle behaviour, with the system seeking to reach an equilibrium that is both unstable and unreachable. Prompted by Pringle's talk, a physical reason for such a double-valued relation was put forward \cite{meyer81}: the vertical structure of accretion discs depends on whether energy transport is largely by radiation or largely by convection, and at temperatures around 10$^4$\,K two solutions are possible -- a convective solution with a low mass flow rate and a radiative solution with a high mass flow rate. The reason for the transition occurring at this temperature is essentially to do with opacity, which drops sharply below 10$^4$\,K because ionization becomes partial as hydrogen recombines.

\begin{figure}[h]
\begin{center}
\includegraphics[width=10cm,angle=0]{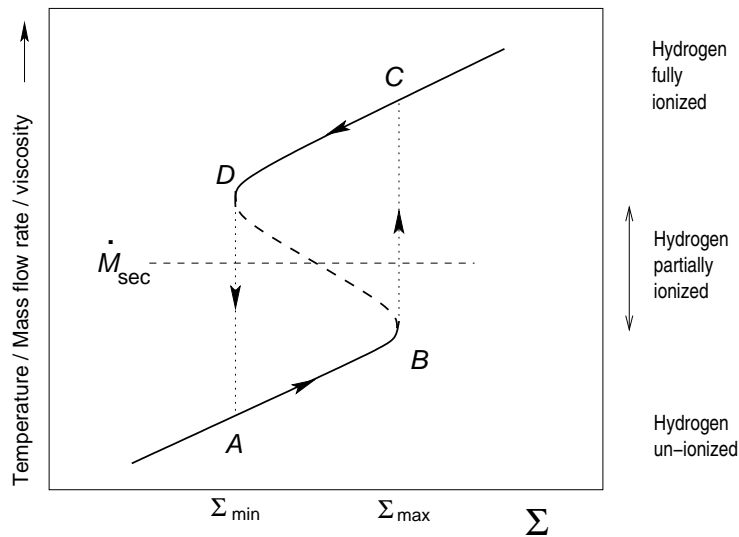}
\end{center}
\caption{\label{f_scurve}Schematic `s-curve' showing the dependence of viscosity on surface density. There is a range of surface density where there are three possible values: two of these, on the solid sections AB and CD, correspond to stable solutions; the section DB is unstable. The low viscosity solution AB corresponds to a cool disc where energy is carried principally by convection, the hydrogen is neutral and the mass flow rate through the disc, $F_{\rm M}$, is low. The high viscosity solution CD corresponds to a hot disc where energy is carried principally by radiation, the hydrogen is ionized and $F_{\rm M}$ is high. Dwarf nova outbursts will occur if the mass transfer rate from the secondary, $\dot{M}_{\rm sec}$, is at a level corrresponding to the unstable branch DB. (Adapted from Fig. 5.7 of \cite{hellier01}.)}
\end{figure}

How does this double-valued relation explain dwarf nova outbursts? Figure \ref{f_scurve} shows the relation, plotted as $\nu \Sigma$ against surface density\footnote{Surface density is the mass per unit area at a point in the disc, defined by $\Sigma = \int_{-\infty}^{+\infty}\rho dz$ where $\rho$ is the mass per unit volume and $z$ is the vertical co-ordinate in the disc.} $\Sigma$. The so-called `s-curve' corresponds to thermal equilibrium, with the heating rate equal to the cooling rate on the curve. Because higher viscosity also means more dissipation in the disc, the vertical axis can also be thought of as either $F_M$ or disc temperature $T$: both increase monotonically with viscosity. From the discussion above, the solid portions of the `s-curve', where viscosity increases with density, are stable, while the dashed section between D and B is unstable -- and indeed there is no equilibrium there. Suppose that $\dot{M}$ has a value that would require $F_M$ to be on the unstable branch if it were to balance $\dot{M}$ (shown as the horizontal line marked $\dot{M}_{\rm sec}$), and consider what happens to a system that starts on the branch AB. Here $F_M$ is low, and in particular $F_M<\dot{M}$. The disc therefore starts to accumulate mass, and so the surface density increases: the system moves along the curve towards point B. When it reaches B, $F_M$ is still less than $\dot{M}$ so the density continues to increase -- but there is no equilibrium curve to follow. Instead, the system moves off into a region where heating exceeds cooling: it becomes thermally unstable, and rapidly moves to the nearest equilibrium at higher $\Sigma$ -- which is on the upper curve, at C. Having reached C, the viscosity, temperature and mass flow rate have all increased and the system now finds itself with $F_M>\dot{M}$. Mass begins to drain out of the disc, and $\Sigma$ decreases again -- so the system now moves down the upper curve, towards D. At D, it again becomes thermally unstable as it tries to decrease $\Sigma$ further, this time entering a region where cooling exceeds heating. There is a rapid transition back to the lower branch, at A, and the cycle starts again. In terms of the dwarf nova outburst, AB corresponds to quiescence, BC to the rise to maximum, CD to a slow decline during maximum and DA to the return to quiescence. Because the viscosity is much higher on the upper branch, by a factor of 100 or so, the viscous timescale for density change is much shorter, so the time taken from C to D is (as observed) much shorter than the period AB in quiescence. The two transitions between the stable branches, BC and DA, occur on a thermal timescale, which for discs is much shorter than the viscous timescale.

These features are qualitatively in agreement with observations, but a detailed match has to take other factors into account (see \cite{cannizzo93} for a review). Most importantly, this discussion was for a particular annulus in the disc. The outburst probably starts in a small region of the disc, but it will then spread through the whole disc as a heating wave. In cases when $\dot{M}$ is low, the outburst tends to start near the centre and propagate outwards (an `inside-out' outburst); when $\dot{M}$ is higher it tends to start near the outer edge and propagate inwards (`outside-in'). The decline from outburst occurs as a cooling wave, which always propagates inwards from the cool outer regions. One problem with the model is that it is not yet possible to predict the value of $\alpha$, so although by making a suitable choice of $\alpha(t)$ it is usually possible to model the observations it is not possible to decide how realistic the model is. Other problems are discussed by Smak \cite{smak00}.

Despite the problems, there are two other significant positive features of the model: firstly, it explains why nova-like systems, with high $\dot{M}$, do not show outbursts -- they are permanently in equilibrium on the upper branch of the s-curve. Secondly, with the addition of another parameter, it provides an explanation for the standstills in Z Cam systems -- these seem to be systems with $\dot{M}$ almost exactly at the level corresponding to point D. Normal outbursts occur when the mass transfer rate is just below that, on the unstable branch. However, if irradiation from the white dwarf and disc causes an increase in $\dot{M}$, it may take the system just above D, onto the stable branch, and outbursts will cease as the system finds an equilibrium. Note that this equilibrium occurs at a point where the system is fainter than at maximum (which would occur at point C), as observed -- Z Cam systems normally `get stuck' about one-third of the way down from maximum.

This disc instability model is currently the canonical picture for dwarf nova outbursts. Historically, it was preceded by a mass transfer burst model, arising from a dynamical instability in the red star; the outburst was taken to be the disc's response to this pulse of enhanced mass transfer. A good summary of that model can be found in the review by Bath \cite{bath85}. Although it is not currently taken seriously, some aspects of the model are relevant to systems where irradiation causes the mass transfer rate to be enhanced (cf. Section \ref{irrad}).

\subsection{SU UMa stars and asymmetric discs}\label{suuma_ob}

\begin{figure}
\includegraphics[width=12cm,angle=180]{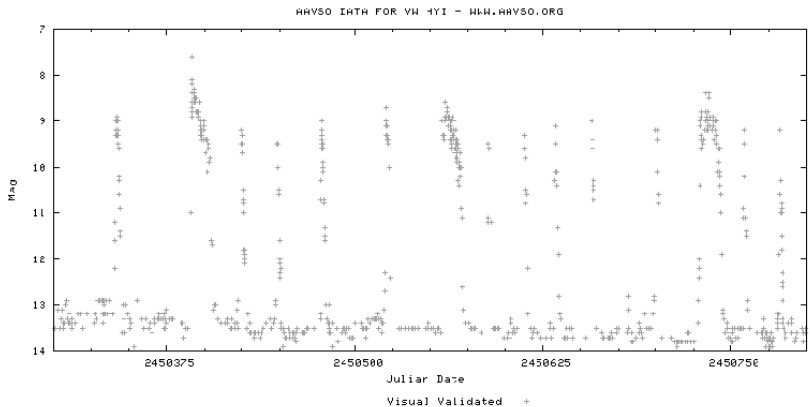}
\caption{\label{f_suuma}A 500-day stretch of the light curve of VW Hydri, a typical SU UMa star, taken from the AAVSO database. Every so often the outburst is larger and longer than normal; it also recurs more regularly.}
\end{figure}

There is a class of dwarf novae, the SU UMa stars, that shows additional features and requires an extra physical mechanism to explain the outbursts. These systems all have rather short periods, but are mainly characterised by the presence of {\it superoutbursts} that occur rather less often than the normal DN outbursts but occur more regularly, last longer and rise to slightly higher luminosities (figure~\ref{f_suuma}). During the superoutbursts, the orbital light curve shows {\it superhumps}\,; these resemble the pre-eclipse humps seen in normal dwarf novae, but they only appear during outburst and they drift in phase relative to the orbital variation. The first superhumps to be observed had periods slightly longer than the orbital period, and gradually drifted to later orbital phases. As more systems were observed, it was found that not all showed these `positive superhumps'; a few show `negative superhumps', whose periods are slightly shorter than the orbital period.

The normal outbursts are explained in the same way as for dwarf novae, except that it seems that the outburst does not leave the disc in exactly the same state as it was at the beginning of the outburst cycle: not all the material stored in the disc has been lost to the white dwarf, so the disc is slightly more massive and larger at the beginning of the next cycle. The disc radius grows slowly from one cycle to the next, until the disc is large enough for tidal forces from the secondary to start distorting it. As shown by Whitehurst in a remarkable thesis that I was fortunate enough to examine (and that was later published: \cite{whitehurst88}, \cite{whitehurst91}), the disc becomes tidally unstable to a 3:1 resonance, in which the particle orbits have frequency $\Omega = 3 \Omega_{\rm orb}$; the resonant radius is within the disc if the mass ratio is $<0.33$. The disc then becomes elliptical, and its long axis precesses. Viscosity is key to this precession: in an inviscid disc, particle orbits in the disc would close after 3 cycles, and the asymmetric structure would rotate at the orbital frequency. The effect of viscosity is that the orbits never close, and the structure drifts slowly relative to the orbital frequency, causing a slow prograde precession. Since the asymmetric structure is also asymmetric in brightness, being brightest where particle orbits intersect, this drift is seen as a slow phase drift of the region of maximum emission, with a period a few per cent longer than the orbital period, as observed for the superhumps.

Why does this happen only during superoutbursts? Osaki (e.g. \cite{osaki96}) argues that it is only when the disc radius is larger than the resonant radius that the tidal effects of the secondary are large enough to remove significant amounts of angular momentum. During a normal outburst, the disc radius is less than the resonant radius, the disc is axisymmetric, little angular momentum is lost, and so not all the mass that has been building up in the disc can be dropped onto the white dwarf. The disc therefore continues to grow slowly in mass, and therefore in radius, during a series of normal outbursts, until finally the disc radius exceeds the resonant radius. The tidal instability then produces an asymmetric disc, the tidal torques are greatly enhanced, and enough angular momentum can be removed from the disc that it can shrink, and lose a lot of mass to the white dwarf. This is the superoutburst, after which the disc is small again and goes back to normal outbursts. Although this model is now widely accepted, and has the virtue of also explaining why the supercycle outbursts occur more regularly, some systems show evidence that enhanced mass transfer may also play a role, possibly driven by irradiation from the disc.

In the last decade it has become apparent that there is a great range of supercycle lengths, from the ER UMa stars with superoutbursts every few tens of days (and very rapidly repeated normal outbursts in between) through the `normal' SU UMa stars with supercycle lengths of a a few hundred days to the very long supercyle systems such as WZ Sagittae, with a period of some 33 years; some of the very long-period systems, such as EG Cancri, show a short series of `echo outbursts' on the decline from superoutburst, but otherwise normal outbursts are rare. The paucity of normal outbursts at both extremes of supercycle length may be related to the fact that both types of system have very short orbital periods, close to the period minimum, and hence very low mass secondary stars.

The difference in supercycle length seems to be related to mass transfer rate, which is largest for the shortest supercycle systems. Interestingly, systems have now been found with mass transfer rates that are too high for outbursts but that nevertheless show superhumps. These {\it permanent superhumpers} are short-period novalikes that satisfy the mass ratio condition needed to produce elliptical discs -- so in fact superhumps don't only occur during superoutbursts: the key ingredient is simply that the disc radius is larger than the resonant radius.

As mentioned earlier, some systems (all nova-likes, with no outbursts) show `negative superhumps', where the superhump period is slightly shorter than the orbital period. The explanation for this is not yet clear, but may involve an accretion disc that is tilted out of the orbital plane; such a disc is expected to precess in a retrograde direction (see \cite{hellier01}, p.92).

Finally, note that tidal torques can have other effects than producing an elliptical disc: they may also generate spiral shocks in the accretion disc. Such shocks first appeared in computer simulations of discs published in 1986, more than a decade before they were observed in the CV IP Peg during an outburst (see reviews by Boffin and Steeghs, \cite{boffin01}, \cite{steeghs01}). The simulations, and the observations, appear to show clear two-armed spirals (figure \ref{f_doppler}), rather like those seen in `grand design' spiral galaxies, although on a very different scale. However, there is not a universal acceptance of the interpretation of these patterns as spiral shocks. First of all, spiral density waves affect the density distribution in the bulk of the disc, but the patterns appear in the observations only in the emission lines, which -- as I explained in Section \ref{spectra} -- arise in the chromospheric outer layers of the disc, either above a natural temperature inversion or in an irradiated region. As Warner has pointed out, this means that it is not actually known whether there is a spiral pattern in the underlying disc. Smak \cite{smak01} has made a similar point, and has suggested an alternative explanation for the patterns seen in Doppler tomograms. He argues that tidal interactions are indeed involved but that they merely cause bunching of orbits, which can produce local thickenings of the disc; these thicker regions stick up into the radiation field from the white dwarf and inner disc and emission lines are therefore produced in their upper layers. His model is able to account for the fact that the two `arms' of the pattern seen in tomograms have different intensities, which would not be expected if they arose from a true two-armed spiral shock.

\section{Secular evolution}

\subsection{The origin of CVs}\label{origin}

The theory of single star evolution (e.g. \cite{iben85}) predicts that the burnt-out core of an intermediate-mass star -- that is, a core that has consumed all its central hydrogen and perhaps also all its central helium -- will contract until it is very dense and the pressure is mainly due to degenerate electrons; the core is then said to be degenerate. The remaining hydrogen and helium, making up the envelope of the star, will expand and the star will go through phases where nuclear fusion occurs in a shell or shells outside the degenerate core. It is observed at these times as a red giant or supergiant, and the properties of the giant are almost entirely determined by the mass of the degenerate core. Eventually, the outer envelope is believed to be lost as a planetary nebula, and the degenerate core is left behind as a hot luminous object that gradually cools and dims to become a white dwarf star, whose composition is either mainly helium or mainly a mixture of carbon and oxygen, depending on mass (Section~\ref{cn_ob}; see also Figure 19 of \cite{iben85}). In either case, the star is now supported against gravity by the pressure of degenerate electrons, with normal gas pressure being important only in a thin surface layer.

This creates an interesting difficulty for CVs, because it means that the white dwarf component must originally have been a red giant star, whose radius would have been several hundred times larger than the current orbital separation from its red dwarf companion. This apparent paradox has led to the current canonical model for the origin of CVs. It is supposed that originally the system consisted of two normal main-sequence stars of different mass, in a large orbit with a period of several years. The larger-mass star would have evolved first and expanded to become a red giant. For a range of initial separations, the red giant star would fill its Roche lobe (Section~\ref{mdot}) and overflow it, transferring mass to its lower-mass companion. As I explained earlier, mass transfer from the more massive to the less massive component leads to orbital shrinkage and hence an increase in the mass overflow rate: the mass transfer is dynamically unstable, so there are significant changes in the mass transfer rate on an orbital timescale. This means that the lower-mass star is receiving mass at a rate much faster than the thermal timescale rate at which it can adjust its structure to accept it, and it rapidly finds itself engulfed in the expanding envelope of the red giant. It is now orbiting in a resistive medium, and the friction causes orbital decay and transfer of angular momentum to the expanding envelope. The details of this {\it common envelope phase} are very complicated and not fully understood (see \cite{iben93} for a good review), but it is generally believed that the net effect is that the lower-mass star and the degenerate core (soon to become a white dwarf) spiral inwards until the envelope has been driven off and the frictional angular momentum loss stops. The predicted end result is a planetary nebula with a close binary at its centre -- and a significant number of planetary nebulae do have binary central stars. Indeed, recent evidence (\cite{demarco06}, \cite{moe06}) suggests that all observed planetary nebulae may arise from binary star systems; planetary nebulae from single stars may simply be too faint to be observed.

The separation of the resulting close binary depends not only on the initial mass ratio and separation but also on the precise details of the common envelope evolution, so it is uncertain, but it is thought unlikely that the orbit shrinks quite far enough for the lower-mass star to end up filling its Roche lobe. More probably, the binary ends up as a white dwarf/red dwarf pair, with an orbital period of a few days; many such systems are known and are now often called `post-common-envelope binaries' or even `pre-CVs'. Some new angular momentum loss mechanism is required to turn these systems into CVs, and I shall discuss possible mechanisms in the next section.

\subsection{Angular momentum loss}\label{aml}

If a binary system is taken to be two point masses in orbit around a common centre of mass, the only angular momentum loss mechanism arises from the emission of gravitational quadrupole radiation. This general relativistic effect depends strongly on the separation of the two stars (inversely as the separation to the fourth power), and is only significant for orbital periods less than about 3 hours. It is therefore important for some short period CVs, as will be seen later, but fails by a large margin to be adequate as a mechanism for turning post-CE binaries into CVs.

To find such a mechanism, it is necessary to drop the point-mass approximation and take into account the properties of the individual stars. The current -- and so far the only serious -- candidate mechanism is known as {\it magnetic braking}. This relies on the red dwarf star having a magnetic field, as is observed to be the case for many single rapidly rotating cool stars, and on the star losing mass via a stellar wind; again, there is evidence for this in many single stars, and the Sun is the closest example of a (weakly) magnetic star that is losing mass. A stellar wind from a rotating star carries off angular momentum as well as mass, but the angular momentum of the material at the surface is not very large because of the relatively small radius of the star. Certainly for the Sun the loss of surface angular momentum is unable to account for its current very slow rotation rate. However, the wind material is hot and ionized and is therefore strongly coupled to the magnetic field lines. The field lines in turn are `frozen in' to the bulk material of the star and therefore corotate with the star. The wind material therefore conserves angular velocity rather than angular momentum as it moves outwards, and it thus gains angular momentum. This process continues until the kinetic energy in the wind exceeds the magnetic energy\footnote{This balance occurs approximately where the velocity of the wind equals the so-called Alfv\'{e}n velocity, $V_A$, and the balance point is known as the Alfv\'{e}n radius. $V_A$ is defined by $V_A = \sqrt{B^2/\mu_0\rho}$, where $B$ is the magnetic field, $\rho$ is the density of the gas and $\mu_0$ is the magnetic constant.}; the wind then breaks free of the magnetic constraint and flows freely outwards, conserving (and carrying away) angular momentum. However, this breaking free occurs only at large radii (around 100 times the photospheric radius for the Sun) and so the angular momentum that is carried away per unit mass is some 10$^4$ times greater than that at the surface. That makes this a very effective angular momentum loss mechanism for single stars, even if the mass loss rate in the wind is quite small.

In close binary systems, such as post-CE binaries, there are significant tidal forces between the stars, which tend both to circularize the orbits and to synchronize the rotation periods of the stars with the orbital period. This tidal interaction has the effect of turning the magnetic braking from an angular momentum loss mechanism for the red dwarf into an angular momentum loss mechanism for the system as a whole: tidal forces spin up the star to keep it in synchronism, and in doing so extract angular momentum from the orbit, turning it into spin angular momentum, which is then carried away by the magnetic wind. This magnetic braking mechanism is believed not only to bring white-dwarf/red-dwarf pairs into contact but to dominate the subsequent secular evolution of the resultant CVs, as will be seen in the next section.

\subsection{The period distribution}\label{pdist}

\begin{figure}
\begin{center}
\includegraphics[width=10cm,angle=270]{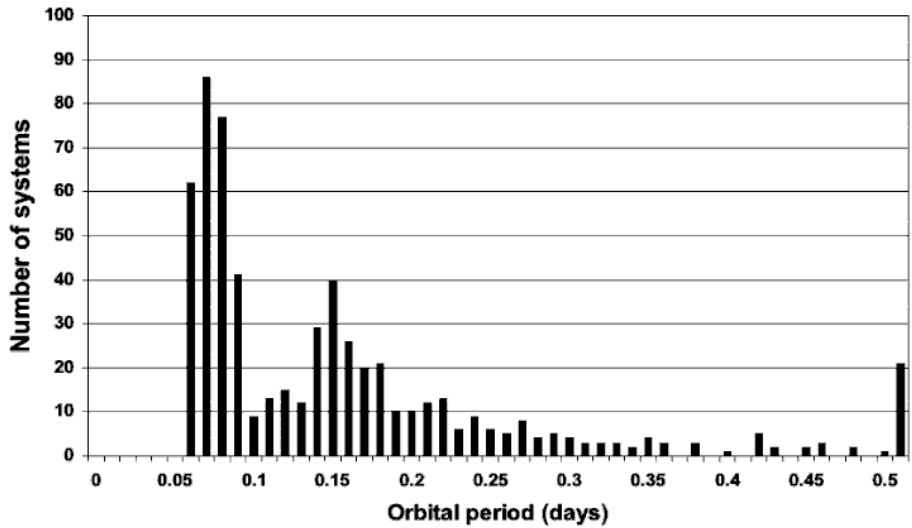}
\end{center}
\caption{\label{f_period}The distribution of orbital periods for CVs, showing the period maximum, period minimum and period gap (data from \cite{ritter06}; the last bin shows all the systems with periods longer than 0.5 days). There is also a small number of systems (not shown) with periods below the period minimum; these are the AM CVn systems (Section~\ref{ac}).}

\end{figure}

It has been known for a long time that there is a very distinctive distribution of orbital periods for CVs (figure~\ref{f_period}). There are almost no CVs with periods longer than about 12 hours, there is a distinct shortage of systems with periods between about 2 and 3 hours\footnote{ Originally, no systems were observed in this period range, so the term `period gap' was introduced; it is no longer quite appropriate, but the term is still used. There remains a clear drop in the number of systems in the 2--3 hour range; this drop is particularly pronounced if one omits the magnetic systems, which may follow a different secular evolution.}, and there is a sharp cutoff at a minimum period of $\sim$77 minutes.

How can these features be explained? The simplest one is the {\it period maximum}. In the canonical model of CVs, the secondary star obeys a main-sequence mass-radius relation and fills its Roche lobe. It can be shown that this leads to a simple relation between the mass of the secondary and the orbital period; roughly, $M_2 \approx 0.1 P_{\rm orb}$, where the mass of the secondary is in solar masses and the orbital period is in hours. In addition, the secondary star is less massive than the white dwarf, which in turn must be below the maximum mass for a white dwarf, the Chandrasekhar mass of 1.44\,M$_\odot$; more realistic models give the maximum mass for a carbon-oxygen white dwarf as about 1.2\,M$_\odot$. Applying these constraints leads to a maximum period of around 12 hours. Some systems are observed with longer periods, so they presumably do not obey all the constraints of the canonical model. In particular, it is believed that the secondary star is probably slightly evolved in these systems, with the system GK Per (orbital period 2 days) being an outstanding example.

The {\it period gap} also has an explanation, but it is not so straightforward and some doubt has been cast on the explanation by the fact that the magnetic systems, which began to be discovered in large numbers by the ROSAT X-ray satellite in the 1990s, do not show the gap nearly as strongly. However, no other plausible explanation has yet been put forward, so I shall describe the model that has emerged from the proposal first put forward by Rob Robinson and colleagues in 1981 (\cite{robinson81}). I first need to describe the general context. As was explained earlier, magnetic braking leads to a steady loss of orbital angular momentum from CVs. This inevitably leads to a decrease in the orbital separation and the orbital period, on a timescale of around 10$^8$ to 10$^9$ years. It is therefore expected that eventually every system will pass through the period range 2--3 hours: so why are few systems seen in that period range? Something must happen to cause most systems to become invisible when they reach an orbital period of around 3 hours, and the obvious explanation is that mass transfer ceases: with no mass transfer, there is (after a brief transition while the disc drains onto the white dwarf) no accretion and no dissipation. The system then becomes much fainter, as is seen, for example, in MV Lyrae from time to time; it was precisely that example that led Robinson et al (\cite{robinson81}) to make their proposal.

The problem with this explanation is to find a mechanism for stopping the mass transfer. The current canonical model attributes this to a change in the internal structure of the star: at a period of about 3 hours, the red dwarf has decreased its mass to the point where, if it were a main sequence star, it would be fully convective. At longer periods, it has a radiative core and a convective envelope, like the Sun. By analogy with the solar dynamo, it is believed that the magnetic field in CV secondaries is generated by a dynamo located in the transition zone between the radiative core and the convective envelope. If the radiative core disappears, so does the transition zone and it is no longer obvious that a dynamo will operate. At the very least, it might be expected to operate in a different mode. It is therefore proposed that, when the secondary star becomes fully convective, the magnetic braking that drives the secular evolution to shorter period suddenly switches off, or is drastically reduced in effectiveness. There will then be a tendency for mass transfer to lead to separation of the two stars (Section~\ref{mdot}) and thus for the mass transfer to cut itself off. Without mass transfer, the system will fade into undetectability.

Another consequence of the stopping of mass transfer is that the secondary star shrinks back inside its Roche lobe: it is out of thermal equilibrium while it is losing mass, and significantly larger than its equilibrium size, so it returns to equilibrium when the mass transfer ceases. However, the evolution to lower period does not stop. Even if magnetic braking has ceased, gravitational radiation will still lead to loss of orbital angular momentum, and it is around a period of 3 hours that that mechanism begins to act on a timescale comparable to the magnetic braking timescale. The invisible system therefore continues to evolve to lower periods, and models show that at a period of around 2 hours the Roche lobe has shrunk sufficiently to bring it back into contact with the now fully convective secondary. Gravitational radiation then becomes the mechanism to drive mass transfer by slowly decreasing the separation between the stars, and the system is re-born as a CV. This is an attractive scenario, but it depends rather crucially on the assumption that magnetic braking stops when the secondary becomes fully convective. This is superficially plausible, but there are no detailed models to support it, and there are examples of low-mass, presumably fully convective, single stars that are magnetically active (cf. \cite{eggleton00}), so there is certainly scope for a completely different model to be put forward; so far, no-one has managed to produce one that both meets all the observational constraints and has achieved universal acceptance.

However, a promising model was put forward by Wu, Wickramasinghe and Warner in 1995 (\cite{warner95i}, \cite{wu95a}, \cite{wu95b}; see also p.478 of \cite{warner95}). This involves irradiation feedback, and does not require any change, dramatic or otherwise, in the magnetic properties of the secondary as the period drops to 3 hours. The model consists of four linked equations that take into account in a self-consistent way both the increase in mass transfer rate caused by irradiation of the secondary and also the effect on the irradiating flux of the consequent larger accretion rate onto the white dwarf. They find that at long orbital periods the secondaries are not very sensitive to irradiation heating and the mass transfer is stable. However, as the period drops below a value in the range 3-4 hours (the precise value depends on the white dwarf temperature), the mass transfer rate as a function of the degree of Roche-lobe filling becomes double-valued, and the system becomes unstable, with a tendency to switch between the two values of mass transfer rate on a timescale of the order of the thermal timescale of the star's envelope, which may be only a few days. A key component of this model, which has not been published in full detail, is that the accretion disc thickens as the mass transfer rate increases, so that the secondary star becomes shielded and the mass transfer rate drops again. The model can explain a number of properties of CVs: (i) the large range of mass transfer rate at a given period, which is especially large in the 3-4 hour period range; (ii) the behaviour of the so-called `anti-dwarf novae': the VY Sculptoris nova-likes, that show sporadic low states, with transitions between states on timescales of days; and (iii) the period gap. The latter is explained by the bistable behaviour in the range 3-4 hours; in the high state, the mass transfer rate is generally rather too high for unstable dwarf nova outbursts, and indeed few dwarf novae are seen in this period range. Similarly, the low state corresponds to the lower stable branch of the s-curve and again few or no DN outbursts are expected. By the time a period of 3 hours has been reached, in the low state, the primary has cooled too far for the system to return to its high state, thus allowing the secondary to relax towards thermal equilibrium; thereafter the evolution through the gap is the same as in the interrupted magnetic braking model.

If either of these pictures is correct, why are some systems seen in the gap? It is possible that some systems first achieve contact only when the period is in the gap. The secondary is then not out of thermal equilibrium, and mass transfer can proceed at the rate imposed by gravitational radiation. In magnetic systems, where the white dwarf has a strong magnetic field, the two stars may be directly coupled by magnetic field lines even once the secondary becomes fully convective, and so accretion can continue -- this probably accounts for the fact that the gap is much less prominent for the magnetic CVs.

What about the third property, the {\it period minimum}? Again, this depends on the changing properties of the secondary star as it loses mass. There is a minimum mass, 0.08\,M$_\odot$, for a normal main-sequence star, below which the centre never becomes hot enough to burn hydrogen. Objects of a lower mass are known as `brown dwarf' stars, and they form by contraction of some initial proto-star until the internal pressure is high enough to stop the contraction. Because of their low central temperature, thermal pressure is insufficient and contraction continues until the central regions are degenerate and the pressure support comes from degenerate electrons, as in white dwarfs. Although the secondaries in CVs start by burning hydrogen in the centre, the burning dies out as the mass decreases and the central temperature drops. What is left as the star approaches the minimum mass limit looks and behaves like a brown dwarf, except that its centre has a slightly higher helium content\footnote{If the CV secondary starts out with a low enough  mass, it will burn hydrogen rather slowly and not much hydrogen will have been consumed by the time nuclear fusion ceases; the centre will then remain H-rich. This is one of the few ways of making H-rich degenerate stars within the age of the Universe.}. In particular, the star changes its mass-radius relation from that for a typical main-sequence star, which is $R \propto M^a$, where $a$ is positive, usually between 0.5 and 1, to that appropriate to a degenerate configuration, which is $R \propto M^{-1/3}$. This qualitatively changes the response of the star to mass loss: instead of shrinking slightly the star now expands. To see how this affects the orbital period, it is helpful (\cite{warner95}, p. 462) to think of mass loss as occurring in two steps: first the mass transfer causes the orbit to expand slightly, breaking contact and cutting off the mass transfer, and then the loss of orbital angular momentum shrinks the orbit again until contact is resumed. While the secondary radius retains its positive dependence on mass, the secondary has shrunk slightly between contact being broken and re-established. Thus the new configuration has a slightly smaller separation and the net effect is to move to shorter period. However, if the radius has a negative dependence on mass, the star actually expands slightly after contact has been broken, so that contact is re-established at a slightly larger separation. The net effect now is to increase the orbital period. Clearly, as the star undergoes what is in fact a smooth transition between a normal main-sequence structure and a degenerate structure, the evolution must change smoothly from period decrease to period increase: there must be a minimum period. This qualitative explanation is borne out by detailed models, although there remains a problem in reproducing the observed value of the minimum period: the theoretical value tends to be significantly smaller (e.g. \cite{barker03}).

Any system that has progressed beyond the period minimum will be very faint: the secondary will have a very small mass and the rate of mass transfer will drop steadily as the orbital period lengthens and gravitational radiation becomes less effective at removing angular momentum. No system in this state has yet been definitely identified. The ultimate fate will be a planetary-size object orbiting a white dwarf.

\begin{figure}
\begin{center}
\includegraphics[width=12cm,angle=270]{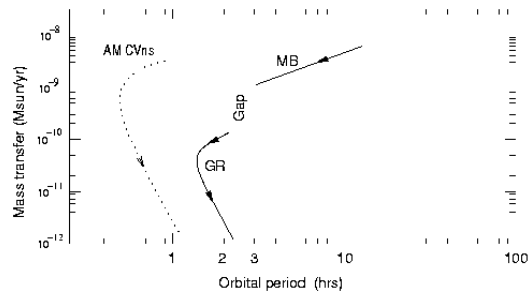}
\end{center}

\caption{\label{f_mdot-p}The evolution of CVs in the mass transfer-period plane. This is controlled by magnetic braking (MB) above the period gap and by gravitational radiation (GR) below the gap; the period gap is shown. At least some of the helium-rich AM CVn systems may follow a parallel path at shorter periods. Based on Figure 4.2 of \cite{hellier01}.}

\end{figure}

Although they are not included in figure~\ref{f_period}, there are actually some systems with periods less than the period minimum. The theory just described makes it clear that these cannot be normal CVs, and in fact they seem (Section~\ref{ac}) to be helium-rich: the AM CVn systems. There are few of these known (only 13 are listed in the January 2006 on-line edition of the Ritter and Kolb catalogue \cite{ritter06}, and two of these, with periods between 5 and 10 minutes, may be something else, as discussed in the review by Nelemans \cite{nelemans05}), but they pose an interesting problem: how did they get to such short periods? The clue is the composition: stars made mostly of helium are generally more compact than hydrogen stars of the same mass, so a white dwarf/helium star binary will come into contact at a rather smaller separation and shorter period than the corresponding CV. It is, in that sense, not surprising that there is a helium sequence to the short-period side of the normal CV distribution and `helium CVs' would be expected to follow the same track in the mass transfer/period diagram as normal CVs, but displaced to shorter period; this is shown in figure~\ref{f_mdot-p}. But why are there helium secondaries in the first place?

There are several possible answers to that question (\cite{nelemans05}), but one possibility is a variant of the common envelope model for hydrogen CVs; with appropriate initial conditions, the initially lower mass star may also evolve to the red giant stage, while the system is still in the detached `pre-CV' phase, and initiate a second common envelope phase that generates a system with a helium secondary star and a white dwarf. The common envelope phase may be enough to bring them into contact, but if not gravitational radiation certainly will, leading to a helium analogue of a normal CV. This picture will produce the track shown in figure~\ref{f_mdot-p}; other scenarios may still produce AM CVn systems, but may not predict a period minimum. There are as yet too few systems to decide whether or not the helium systems show a period minimum.

\subsection{Population synthesis}\label{popsyn}

As well as explaining the various gaps in the period distribution, there have been attempts (\cite{politano96}, \cite{dekool92}) to calculate the relative numbers of CVs that are expected in each period range. This requires choosing starting conditions that are representative of observed single stars and following the evolution of a large number of pairs of such stars through the common envelope stage and into contact. The initial conditions include such things as the distribution of single stars with mass and the distribution of mass ratios and orbital parameters in binary systems, and a starting distribution of binaries is chosen by picking parameters randomly from these distributions, for example using a Monte Carlo scheme \cite{dekool92}. In order to gain a statistically significant result, a very large population of binaries has to be followed, which requires a fast evolutionary algorithm -- this is normally obtained by creating a semi-analytical fit to a grid of full stellar evolution calculations. Other factors that need to be considered in the calculations are how to treat the common envelope phase, how to represent the angular momentum loss mechanism and how to calculate mass transfer and its effects on the two stars. There are three stages: calculation of the formation rate of CVs, with the attendant uncertainties of the common envelope phase, calculation of their evolution with period, taking account of various angular momentum loss mechanisms, and then comparison with observations, after allowing for the very severe observational selection effects (for example, longer-period systems with higher-mass secondaries are more likely to be recognised as CVs; at the other extreme, post-period-minimum systems are very hard to detect). Despite all these uncertainties, there is reasonably good qualitative agreement between the predictions and the observations, especially for the overall space density of CVs, although it is hard to obtain a quantitative match with all the details. For example, the recent realisation that many CVs have somewhat evolved secondaries is in conflict with early predictions that less than 10\%\ of CVs should form with evolved donor stars. Attempts to modify the predictions by including a thermal-timescale mass transfer phase (e.g. \cite{kolb05}) have been only partially successful.

\subsection{Irradiation and hibernation}\label{irrad}

One of the puzzling features of CVs is that there is a great range of mass transfer rates at a given orbital period: WZ Sge systems co-exist with permanent superhumpers below the period gap, while normal dwarf novae also show a range of mass transfer rate (as evidenced by their disc brightness) at a given period, going right up to the high rates seen in old novae and nova-likes (cf. Fig. 9.8 in \cite{warner95}). Given the model of secular evolution that underlies our understanding of the period distribution, it is not clear why there is not a single mass transfer rate at each period. What additional parameter could cause the spread? One that has been suggested on a number of occasions is irradiation of the secondary star by the accretion flow and the white dwarf. There is plenty of observational evidence that irradiation occurs, and it is argued that it may cause an increase in mass transfer (cf. the explanation in Section \ref{dn_ob} for the standstills in Z Cam systems and that in Section \ref{pdist} by Wu et al. for the period gap).

One proposal has been that all systems undergo some sort of thermal mass-transfer cycle, with mass transfer causing irradiation that then increases the mass transfer rate until the star has had time to adjust its thermal structure and contract, cutting the mass transfer rate again. Detailed calculations (e.g. \cite{buning04}) show that such thermal cycles are possible, but under rather restrictive conditions that make it unlikely that these cycles can provide a general mechanism for explaining the observed range of mass transfer rates. One problem with this and related models by Kolb, King and others (e.g. \cite{king95}) is that the only thermal timescale in the model is that of the whole star, which is too long to explain some of the shorter timescale variations that are seen (note that the model of Wu et al. \cite{wu95b} has no such restriction).

Despite the problems with some thermal cycle models, irradiation probably does play a significant role in accounting for the large range of mass transfer rate. The largest mass transfer rates observed are around $2\times10^{-8}$\,M$_\odot$\,yr$^{-1}$, which, at least for periods less than 6\,h, is higher than can be maintained by magnetic braking, and the high mass transfer systems tend to be those with the hottest white dwarfs and boundary layers and thus the highest irradiating fluxes. The high temperatures (in excess of $5\times10^4$\,K) are themselves generated by the high mass transfer rates, but thermal runaway is averted by the fact that the disc tends to swell up when the mass flow rate is high, and this shields the secondary, limiting the effects of the irradiation \cite{wu95b}.

Why should the temperatures be high enough in the first place to start the irradiation process? An initially attractive proposal that also unifies the various types of CVs is the hibernation scenario proposed by Shara and colleagues and usefully summarised in section V of his 1989 review \cite{shara89}. The original motivation was the observation that classical novae seemed to be rarer by about two orders of magnitude than expected from models of the outbursts and from the outburst frequency deduced from observations of the galaxy M31. In outburst, novae reach a luminosity many times larger than the normal luminosity of the red dwarf, so irradiation from the explosion on the white dwarf might be expected to have a dramatic effect, leading to a significant increase in the mass transfer rate. The idea was that as the outburst died away the effects of the irradiation would persist for a few centuries (in line with observations of historical novae), but that after that the mass transfer rate would fade away, with the two stars possibly losing contact altogether, and the system would become essentially invisible -- it would be in hibernation, accounting for the apparent shortage of old novae. This low state would persist for most of the time before the next outburst, but over the century or two before that the system would come back into contact and the mass transfer rate would build up again. During the intermediate phases of moderate mass transfer, both fading and recovering, the system would be seen as a dwarf nova. This picture neatly ties together old novae, novalikes and dwarf novae as different stages in the life of a single object, and indeed some old novae have been seen to show dwarf nova style outbursts, so at least part of the idea is probably correct.

However, the proposal proved controversial, and it has led to a variety of papers both in favour of it (e.g. \cite{duerbeck92}) and against it (e.g. \cite{naylor92}); see also section 9.4.3 of Warner's book \cite{warner95}. The key problem with the hibernation picture is to explain why the mass transfer rates drop so low that the systems become essentially unobservable; a recent study of relative numbers of novalikes and dwarf novae \cite{martin05} suggests that hibernation is only viable if the rate of magnetic braking is an order of magnitude less than normally assumed. There also seems no obvious mechanism that would make all old novae break contact, with detailed calculations \cite{livio91} suggesting that for short period systems the frictional angular momentum loss as the nova shell engulfs the secondary causes the two stars to come closer together.

\section{What do CVs teach us?}

\subsection{Insights into accretion}\label{accretion}

The theory of spherical accretion was worked out by Bondi, Hoyle and Lyttleton (\cite{hoyle39}, \cite{bondi44}; see also \cite{edgar04}) at a time when it was thought that accretion of material from the interstellar medium was a key component in explaining properties of the Sun. Although that is no longer believed, the theory of accretion was revived around 1970 (\cite{lynden-bell69}, \cite{shakura73}) when it was realised that quasars and other active galactic nuclei (AGN) were probably powered by accretion onto a central black hole, although via a disc rather than by a spherical inflow. It soon became apparent that the observational study of accretion processes could be carried out more effectively in the much closer systems that are now known as X-ray binaries and cataclysmic variables -- accretion is a key ingredient in their structure and in the structure of other interacting binaries, and their study can be used to inform ideas about the much more energetic events happening in AGN. An excellent monograph by Frank, King and Raine (\cite{frank02}) lays out the theory of accretion and its application to binaries and to AGN.

An example of an insight into the properties of accretion discs gained by studying CVs is the discovery of spiral structure during outburst in the disc of IP Pegasi and of some other CVs, using the technique of Doppler tomography \cite{steeghs01}. The implications for angular momentum transport and viscosity in the disc are still being worked out, but it seems likely that the spiral arms are generated by the tidal effect of the companion (\cite{boffin01}; but see Section \ref{suuma_ob}). If so, extension of this theory to accretion discs around AGN is unlikely unless some other mechanism can mimic the effect of tides. However, there could be implications for proto-planetary discs around young stars if the star forms in a binary or in a dense cluster with many nearby stars to perturb the disc.

Another effect of tides is the formation of an elliptical disc under some circumstances (Section~\ref{suuma_ob}), although again an extension to AGN is unlikely, unless AGN discs possess some non-axisymmetric instability that causes them to precess like the elliptical discs in SU UMa systems.

Perhaps more generally, a comparison of the hydrogen discs in most CVs with the helium discs in AM CVn stars will throw light on the effect of composition on the structure of accretion flows, which may be relevant to AGN. However, the details of accretion are often rather different in AGN discs, and the main gain for AGN studies has been the development of the general theory of disc accretion and its testing by observations of interacting binaries.

\subsection{Tests of stellar evolution}\label{evolution}

Just as variable stars provide a formidable challenge for theories of stellar structure, because they test the time dependence as well as the steady state, so interacting binary stars provide tests of stellar evolution that cannot be made from isolated single stars. Many of these tests, such as the apsidal motion test that probes the internal structure of stars, and the effects of mass transfer on both the mass-losing and the mass-gaining star, are not exclusive to CVs (and indeed the essentially circular orbits of CV components rules out any application of the apsidal motion test, which looks at how fast the major axis of the orbital ellipse precesses). However, the secular evolution of CVs provides some interesting variations on normal stellar evolution. First of all, the initial Roche lobe overflow that forms the common envelope (Section~\ref{origin}) cuts off the normal evolution of the giant star and prematurely produces a white dwarf remnant -- so this provides a route to lower-mass white dwarfs than could have been produced by single star evolution within the life of the Universe.

Secondly, two successive common envelope phases can produce a system of two helium stars (Section~\ref{pdist}), whereas it is very difficult to produce a helium star at all by single star evolution within the age of the Universe. This allows tests of the effects of composition differences that would otherwise be impossible.

Thirdly, the response of a star to mass loss can be theoretically predicted, but only in semi-detached binary systems can the theory be tested; CVs provide a good example, although the theory was first used to explain Algol systems, which consist of a red sub-giant star losing mass to a more massive unevolved companion. Algols are in some sense failed CVs: they also underwent a period of rapid mass transfer from the initially more massive star to its companion, but the effect was simply to reverse the mass ratio and maintain contact rather than to produce a common envelope and a more compact system.

\subsection{Tests of dynamo theory}\label{dynamos}

As well as the strong magnetic fields seen in sunspots and prominences, the Sun has a weak general field that is roughly dipolar and reverses every 11 years or so (although the field structure is not as simple as that sounds -- there is a tendency (e.g. \cite{webb84}) for the field to reverse at one pole anything from 6 months to 18 months before it does so at the other!). Models of the solar field are based on the assumption that it is produced by an oscillatory but not exactly periodic dynamo with a characteristic timescale of around 11 years, and it is generally believed that the seat of the dynamo is at the transition layer between the radiative core and the convective envelope and that differential rotation plays a vital role in the dynamo mechanism. The pattern of solar activity (sunspots, prominences, flares, coronal mass ejections etc.) varies with the magnetic cycle and similar activity cycles are observed in many other cool stars. Differential rotation has been detected in some of them (\cite{cameron01} and other chapters in \cite{boffinetal01}). It is therefore generally believed that dynamos operate in those stars as well. The activity is strongest in rapidly rotating stars, as for example in the RS CVn binaries, which are detached but very close systems with orbital periods of a few days to a few tens of days. Doppler tomography and Zeeman tomography have revealed active regions on the surfaces of some of these stars (\cite{donati01}, and other chapters in \cite{boffinetal01}, especially for studies of HR 1099).

Despite all this evidence, there is still no completely satisfactory model for a stellar dynamo, even for the Sun. The very rapid rotation of the secondaries in CVs, with rotation periods of only a few hours, allows the possibility of testing dynamo theories in the most extreme parameter ranges available to date. The advent of Roche tomography (Section~\ref{tomography}) has revealed spots on
some of these secondaries and the comparison of the spot patterns with those on less rapidly rotating single stars will provide new tests of theory. Observation of the magnetic activity in these close binaries will also allow tests of the effects of tides on dynamos and field structures -- if dynamos require the presence of differential rotation to drive them, what will tidal synchronisation do to them?

\section{Conclusions}\label{conc}

Studies of CVs have come a long way since the realisation of their binary nature in the 1950s, and it is quite remarkable that it is now possible to produce a map of starspots on the secondary (figure \ref{f_aeaqr}). With more than 600 systems now known, and many related objects, there is a sound statistical basis for discussing their long-term evolution. Nonetheless, there are still many unsolved problems and intriguing puzzles. Most people believe that the source of the viscosity in the disc is turbulence induced by a magneto-rotational instability (MRI), but it is not yet possible reliably to calculate the value of the viscosity parameter $\alpha$. However, the consequences of the MRI are now being explored in great numerical detail and Balbus argues \cite{balbus03} that the use of a single viscosity parameter is in any case no longer appropriate because of the richness and subtlety of the magnetic effects. Nova outbursts are understood in principle, but mass accretion at the rate seen in pre-novae would not produce explosions. The limit-cycle model for dwarf nova outbursts is generally accepted, but there are still uncertainties about how to explain all the details, especially for the SU UMa stars. There seems little doubt that the secondary stars in CVs are magnetically active, but how do they maintain a dynamo in the presence of tidal forces? Is there differential rotation? If not, do our current theories of dynamos in other stars need to be re-thought?

This article has tried to give the flavour of the main issues in CV research, but it has only scratched the surface, and given a few of the more established results; there are plenty of problems left for the future.

\section*{Acknowledgments}

I am grateful to Brian Warner and Coel Hellier for comments and suggestions, and to John Cannizzo, Vik Dhillon and Ren\'{e} Rutten, Coel Hellier, and Chris Watson for providing figures \ref{f_sscyg}, \ref{f_roche1}, \ref{f_scurve} and \ref{f_mdot-p}, and \ref{f_aeaqr} respectively. Figure~\ref{f_suuma} is based on visual data supplied to the AAVSO database by the Royal Astronomical Society of New Zealand (RASNZ), Variable Star Section, while figure \ref{f_period} has been constructed from data extracted from the {\it Vizier} version of the Ritter and Kolb catalogue \cite{ritter06}, held at the CDS, Strasbourg. Figure \ref{f_cv} is \copyright\ CUP; figure \ref{f_hump} is \copyright\ the editors of {\it Astronomy and Astrophysics}\,; figures \ref{f_zcha}, \ref{f_doppler} and \ref{f_aeaqr} are \copyright\ RAS.

\section*{References}

\bibliography{2006review,iaujrnls}

\begin{quote}
\footnotesize The author was educated in Glasgow, but has worked at Sussex since 1968. He began his research life as a theoretical astronomer, working on fluid motions within rotating stars and binary stars, but in the mid-1980s he was seduced by the attractions of observational work, and has since then pursued the properties of the cool component in CVs. His group was amongst the first to produce maps of the surfaces of some of these stars, and his current research student is mapping starspots on them.
\end{quote}

\end{document}